\documentclass[pre,amsmath,amssymb,twocolumn,nofootinbib,floatfix]{revtex4-2}
\usepackage[colorlinks=true,linkcolor=blue,urlcolor=blue,citecolor=blue]{hyperref}
\usepackage{graphicx}
\usepackage{bm,color}
\usepackage{times}
\usepackage{color}

\renewcommand{\doi}[2]{\href{http://dx.doi.org/#1}{#2}}
\newcommand{\arxiv}[1]{\href{http://arxiv.org/abs/#1}{#1}}
\newcommand{\link}[2]{\href{http://#1}{#2}}

\newcommand{\Eq}[1]{Eq.~(\ref{#1})}
\newcommand{\Eqs}[1]{Eqs.~(\ref{#1})}
\newcommand{\eq}[1]{(\ref{#1})}
\newcommand{\half}{\frac12}
\newcommand{\bea}{\begin{eqnarray}}
\newcommand{\eea}{\end{eqnarray}}
\newcommand{\rme}{\mathrm{e}}
\newcommand{\rmd}{\mathrm{d}}
\newcommand{\nn}{\nonumber}
\renewcommand{\epsilon}{\varepsilon}
\newcommand{\ca}[1]{{\cal #1}}
\newcommand{\be}{\begin{equation}}
\newcommand{\ee}{\end{equation}}
\newcommand{\Fig}[1]{\includegraphics[width=\columnwidth]{./#1}} 
\newcommand{\fig}[2]{\includegraphics[width=#1\columnwidth]{./#2}}

\graphicspath{{figures/},{}}

\renewcommand{\log}{\ln}

\begin{document}

\title{Mean-Field Theories for Depinning and their Experimental Signatures}
\author{Cathelijne ter Burg and Kay J\"org Wiese}
  \affiliation{\mbox{Laboratoire de Physique de l’E\'cole Normale Sup\'erieure, ENS, Universit\'e PSL, CNRS, Sorbonne Universit\'e,} \mbox{Universit\'e Paris-Diderot, Sorbonne Paris Cit\'e, 24 rue Lhomond, 75005 Paris, France.}}

\begin{abstract}

Mean-field theory is an approximation replacing an extended system by a few   variables. For depinning of elastic manifolds, these are the position of its center of mass $u$, and the statistics of the  forces $F(u)$. There are two proposals how to model    the latter: as a random walk (ABBM model), or as uncorrelated forces at integer $u$ (discretized particle model, DPM). While for many experiments the ABBM model (in the literature misleadingly equated with mean-field theory)   makes quantitatively correct predictions for the distributions of velocities, or avalanche size and duration,  the microscopic disorder   force-force correlations cannot grow linearly, and thus unboundedly as a random walk,  with distance. Even the  effective (renormalized)
disorder forces which do so at small distances are bounded at large distances. To describe both regimes, we model forces as an Ornstein Uhlenbeck process. The latter has the statistics of a random walk at small scales, and  is uncorrelated at large scales.
By connecting to  results    in both limits, we solve the model  largely analytically, allowing us to describe in all regimes the distributions of velocity, avalanche size and duration. 
To establish experimental signatures of this transition, we study  the response function, and the correlation function of position $u$, velocity $\dot u$ and forces $F$ under slow driving with velocity $v>0$. 
While at $v=0$ force or position correlations have a cusp at the origin and then  decay at least exponentially fast to zero, this cusp is rounded at a finite driving velocity. 
We give a detailed analytic analysis  for this rounding by   velocity, which allows us, given experimental data, to extract the time-scale of the response function, and to reconstruct the force-force correlator at 
$v=0$. The latter is the central object of the  field theory, and as such contains detailed information about the universality class in question. We test   our predictions by careful numerical simulations extending over up to ten orders in magnitude. 
\end{abstract} 

\maketitle

\section{Introduction}

\subsection{Mean-field theories}
The framework of disordered elastic manifolds covers such diverse systems as contact-line depinning  \cite{LeDoussalWieseMoulinetRolley2009}, charge-density waves,  magnetic domain walls \cite{UrbachMadisonMarkert1995,DurinZapperi2000,KimChoeShin2003,DurinZapperi2006b,LemerleFerreChappertMathetGiamarchiLeDoussal1998}, earthquakes \cite{JaglaKolton2009,Kagan2002,DSFisher1998,FisherDahmenRamanathanBenZion1997,PaczuskiBoettcher1996,BenZionRice1993,CrisantiJensenVulpianiPaladin1992,BurridgeKnopoff1967,GutenbergRichter1956,GutenbergRichter1944}, shear of micro-pillars \cite{CsikorMotzWeygandZaiserZapperi2007} and stretching of a knit \cite{PoinclouxAdda-BediaLechenault2018}. Many of these experiments, or at least aspects thereof, are successfully described by {\em mean-field theory}. But what exactly is meant by {\em mean-field theory}? Let us define {\em 
mean-field theory as a theory which reduces an extended system to a a single or few  degrees of freedom}. For depinning these are  $u$, the center of mass of the interface, and the correlations of $F(u)$, the forces acting on it. The center of mass of the interface follows the  equation of motion
\be\label{EOM}
 \partial_t   u(t) = m^2 \left[ w-   u(t) \right] +     F \big(u (t)\big) .
\ee
The first term is the force exerted by a confining well, equivalent to a Hookean spring with spring constant $m^2$.  
The second term $F(u)$ is a random force, possibly the derivative of a random potential, $F(u) = -V'(u)$.
Specifying the correlations of $F(u)$  specifies the system, and    {\em selects one   mean-field theory}. However, when the reader   encounters the term  ``mean-field theory''   in the literature, it is quite generally employed for a model where the forces perform a random walk, 
\bea
\label{219}
 \partial_u F \big(u  \big) &=&   \xi(u) ,  \\
 \label{220}
  \left< \xi(u)\xi(u') \right> &=& 2   \delta (u-u').
\end{eqnarray}
This model was introduced in 1990 by Alessandro, Beatrice, Bertotti and Montorsi (ABBM) \cite{AlessandroBeatriceBertottiMontorsi1990,AlessandroBeatriceBertottiMontorsi1990b} to describe magnetic domain walls, and is nowadays referred to as the ABBM model. 

The {\em forces} $F(u)$ are the {\em coercive magnetic fields}  pinning the domain wall, which were observed experimentally  to change with a  {\em seemingly} uncorrelated function $\xi(u)$ \cite{VergneCotillardPorteseil1981}. 
The decision of ABBM \cite{AlessandroBeatriceBertottiMontorsi1990} to model $\xi(u)$  in \Eq{220} as a white noise  is a  strong assumption, a posteriori justified by the applicability to experiments \cite{AlessandroBeatriceBertottiMontorsi1990b}. 
It means that $F(u)$ has the statistics of  a random walk, with  force-force correlations   
\be\label{DeltaABBM} 
\Delta_0^{\rm RW}(0) - \Delta_0^{\rm RW}(u-u') :=  
\half \overline{ \left[ F(u)-F(u')\right]^2} =   |u-u'| .
\ee
Actually, \Eqs{219}-\eq{220} is the final model analyzed by ABBM.
What they  considered first are forces modeled as  an Ornstein-Uhlenbeck process
%
%
%
\be\label{Ornstein-Uhlenbeck}
 \partial_u F \big(u  \big) =  -   F(u)+ \xi(u).
\ee
Note that by rescaling $m^2$ and time $t$ in \Eq{EOM}, noise strength ``2'' in \Eq{220} and the   amplitude ``1'' for the restoring force in \Eq{Ornstein-Uhlenbeck} can always be achieved. Thus the only parameter of relevance is $m^2$. 
With  the    noise     in \Eq{220}, the process \eq{Ornstein-Uhlenbeck} has correlations 
(see appendix \ref{a:OU-correlations}), 
\be\label{FFOU}
\Delta_0^{\rm OU}(u-u') := \overline {F(u) F(u')}^{\rm c} =  \rme^{-|u-u'|},
\ee
i.e.\ it is uncorrelated at large distances.
One explicitly checks that the small-distance behavior of $\Delta_0^{\rm OU}(u-u')$ is as in \Eq{DeltaABBM}. 
This is the model we study in this article. Contrary to the claim made by ABBM in Ref.~\cite{AlessandroBeatriceBertottiMontorsi1990} (beginning of section III), even at low domain-wall velocities the reduced model \eq{219} is only valid at small scales, and there is always an observable, namely the renormalized disorder correlator defined below in \Eq{16}, which quantifies whether forces are distributed according to an Ornstein-Uhlenbeck process, or a random walk. 
\begin{figure}
\Fig{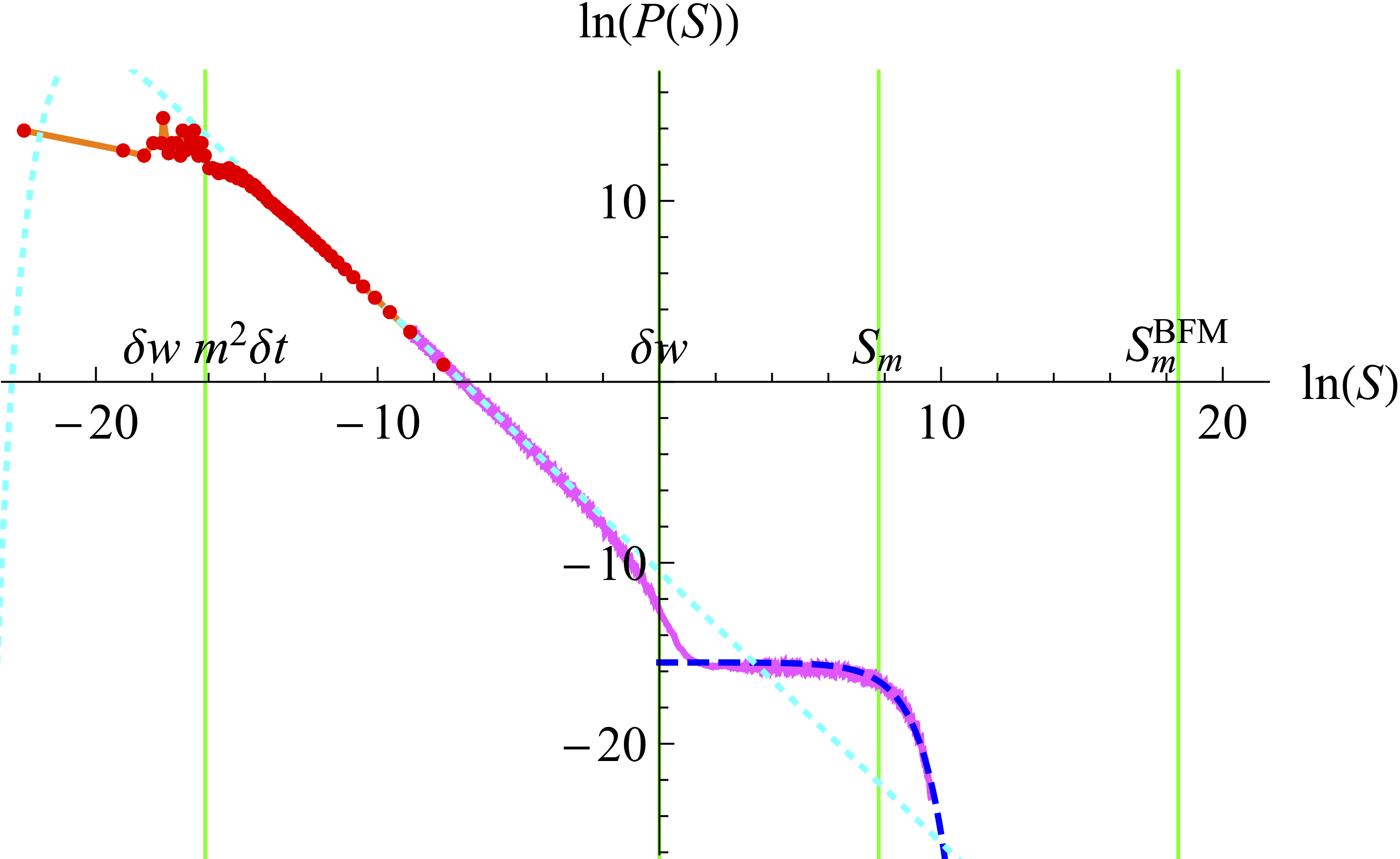}
\caption{Avalanche-size distribution $P(S)$ for a particle  with forces given by  \Eq{Ornstein-Uhlenbeck}.  The theoretical curves are  the  kicked ABBM model as given by \Eq{PwS(S)=ABBM} (cyan dotted), and the DPM
as given by \Eq{PofS-discrete} (blue, dashed). INS,  $m^2=10^{-4}$, $\delta w=1$, $\delta t=10^{-4}$, $S_m:=\left<S^2\right>/(2\left< S\right>)=2408.89$, $\rho_m=2329.95$,   $N=10^8$.}
\label{f:1}
\end{figure}

An example for the differences between the two models is given on  figure \ref{f:1}, which shows the avalanche-size distribution, assuming forces generated by the Ornstein-Uhlenbeck process  \eq{Ornstein-Uhlenbeck}.   
%
%
One sees that for small avalanche sizes $S$, the probability distribution follows $P(S)$ as predicted for the ABBM model, with no adjustable scale, and a critical exponent $\tau_{\rm ABBM}= 3/2$ (defined in table \ref{t:compare}). However, when the avalanche size reaches the correlation length of the forces, which according to \Eq{FFOU} is $\xi_{F}=1$ in our units, the avalanche-size distribution crosses over to a pure exponential, formally equivalent to an avalanche-size exponent $\tau_{\rm DPM}=0$. As can be seen on Fig. \ref{f:1} and   summarized in table \ref{t:compare}, it also drastically changes the scaling of  the large-scale cutoff, from $S_m^{\rm ABBM} \equiv S_m^{\rm BFM} \sim m^{-4}$, to $S_m^{\rm DPM}\sim m^{-2}$.

Up to now, we only discussed mean-field models. 
Field theory \cite{LeDoussalWieseChauve2003,LeDoussalWieseChauve2002,ChauveLeDoussalWiese2000a}  gives a more differentiated view: First of all, mean-field theory should be applicable for $d=d_{\rm c}$ \cite{FedorenkoStepanow2002,LeDoussalWiese2003a},  a case which contains magnets with strong dipolar interactions \cite{DurinZapperi2006b}, earthquakes \cite{DSFisher1998}, and micro-pillar shear experiments  \cite{CsikorMotzWeygandZaiserZapperi2007}. 
As $F(u)$ has the statistics of a random walk, the  (microscopic) force-force correlator of \Eqs{219}-\eq{220}  as  given in \Eq{DeltaABBM},   
grows linearly with distance. 
A linearly increasing correlation function is  at the microscopic level    predicted for the correlations of the potential,  $R(0)-R(u):= \half \left <  [ V(u){-}V(0) ]^2\right> $, if the disorder is of the random-field type, the strongest microscopic disorder at our disposal \cite{WieseLeDoussal2006,WieseRPP}.  We know of no microscopic mechanism to generate the correlations of \Eq{DeltaABBM}. 
On the other hand, the effective (renormalized) force-force correlator $\Delta(u)$ has a cusp \cite{MiddletonLeDoussalWiese2006,RossoLeDoussalWiese2006a,WieseRPP}, so \Eq{DeltaABBM} with $ |\Delta'(0^+)|=1$ is an approximation, valid for small $u$.  The ABBM model defined by  \Eqs{EOM}-\eq{220} should then be viewed as  an effective  theory, arriving {\em  after renormalization}, and valid for small $u$ only. 

If indeed the microscopic disorder has the statistics of a random walk, then the force-force correlator \eq{DeltaABBM} does not change under renormalization, as is easily checked by inserting it into the 1-loop  or 2-loop   flow equations for depinning \cite{LeDoussalWieseChauve2002,ChauveLeDoussalWiese2000a}. Counting of derivatives for higher-order corrections proves that this statement persists to all orders in perturbation theory. Even an extended (non-MF) system where each degree of freedom sees a   force which has the statistics of a random walk, the Brownian-force model (BFM) introduced in Ref.~\cite{LeDoussalWiese2012a}, is stable under renormalization, and has a  a roughness exponent 
$
\zeta_{\rm BFM} =  4-d
$, where $d$ is the dimension of the elastic object. This was indirectly    verified numerically in  Ref.~\cite{ZhuWiese2017}.

Our discussion below shows that the ABBM model \eq{219}-\eq{220} is   adequate only at small distances, but   fails at larger ones, where the force-force correlator decorrelates. We therefore expect that at large distances it crosses over to 
a model of uncorrelated random forces. 
Such a model,  which we term the {\em discretized particle model} (DPM), was introduced in Ref.~\cite{LeDoussalWiese2008a}. 
\begin{figure}
\Fig{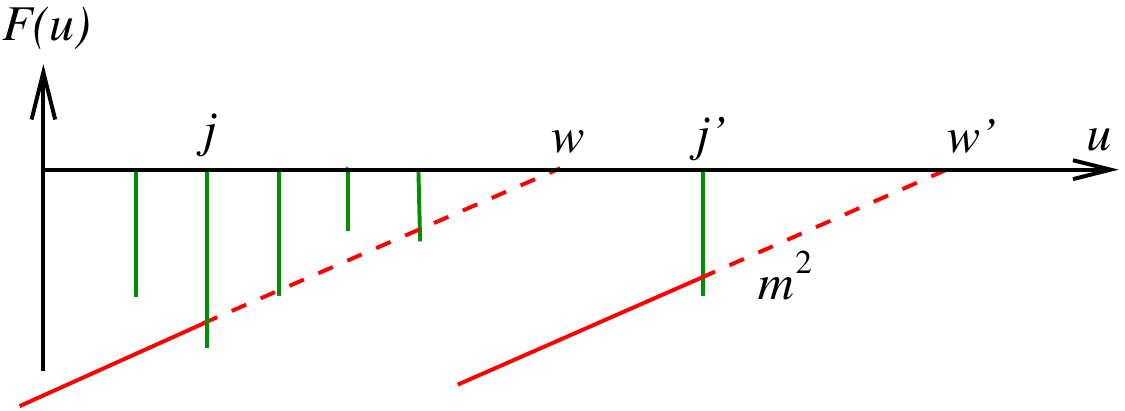}
\caption{Geometric solution to find $u_w$ as a function of $w$ for the DPM model of \protect\cite{LeDoussalWiese2008a}.}
\label{f:DPM}
\end{figure}
In this model, the random forces are modeled by needles at integer positions, blocking the advance of the particle.
Figure \ref{f:DPM} shows this model, and how to geometrically construct the solution of \Eq{EOM}: Draw a straight line $f(u) := m^2 (u-w)$ of slope $m^2$, intersecting the horizontal axis ($F=0$) at $u=w$. As long as $f(u)<F(u)$, the particle advances. The motion is stopped at the first obstacle this line encounters. On  Fig.~\ref{f:DPM}, this is $u(w)=j$ and $u(w')=j'$. Remains to specify the correlations of the random forces $F(j)$, $j \in \mathbb{N}$. As field theory in general supposes Gaussian disorder, we choose   $F(j)$ to be Gaussian distributed, with variance
\be\label{DPMFF}
\overline {F(j) F(j')} =    \delta_{j,j'}.
\ee
In this article, we wish to study the crossover from ABBM-disorder   given by \Eqs{219}-\eq{DeltaABBM}, to the correlations \eq{DPMFF}. 
We do this by analyzing the Ornstein-Uhlenbeck process (\ref{Ornstein-Uhlenbeck}). 

To compare Ornstein-Uhlenbeck forces with the DPM, let us consider their microscopic disorder force-force correlator. 
For the DPM, with forces  constant between integers\footnote{To render 
$\overline{F(u) F(u')}$   function of $u-u'$ only,   the needle  forces $F(j)$, $j \in \mathbb N$ of Fig.~\ref{f:DPM} are extended to $F(u)=F(j)$, $j\le u<j+1$. The solutions $u_w$ of   Fig.~\ref{f:DPM}  change, $u_w \to u_w-\delta$, with $0\le \delta<1$, negligible for small $m^2$.}, it reads 
\be\label{partF-cor}
\Delta_0^{\rm DPM}(u-u')= \overline{F(u) F(u')}^{\rm c} = {\rm max}(1-|u-u'|,0)\ .
\ee
Both models have a linear (microscopic) cusp, with 
\be
- \Delta_0'(0^+) = 1\ .
\ee
On the other hand, the integral over their force-force correlations is different,  
\bea\label{FF-OU}
\int_{{-\infty}}^{\infty} \rmd u\,  \Delta_0^{\rm DPM}(u) &=&  1,
\\
\label{FF-particle}
\int_{{-\infty}}^{\infty} \rmd u \, \Delta_0^{\rm OU}( u) &=&    2.
\eea
Our working hypothesis to be checked below is that the two models have the same universal large-scale properties, with the same scale, and without any additional parameter.
Nevertheless, we expect that non-universal quantities such  as the critical force  might be shifted. 
 
\Eq{Ornstein-Uhlenbeck} also serves as an  effective theory for the crossover observed in systems of linear size $L$, from a regime with $mL\gg 1$ described by an extended elastic manifold,  to a single-particle regime   described by the DPM. This crossover has indeed be seen in numerical simulations for a line with periodic disorder \cite{BustingorryKoltonGiamarchi2010}.

\subsection{The effective disorder and measurements}
\subsubsection{Force correlator at finite driving velocity}
\label{s:The effective disorder, and  rounding of the cusp by a finite driving velocity}

The question we are now turning to is the following: What can   experiments teach us about the underlying field theory?  Can the crossover be seen in an experiment?

Suppose the system is driven quasi-statically, i.e.~
\be
w = v t,
\ee
and we wish to study the limit of $v\to 0$. To do so, parameterize the solution of \Eq{EOM} as 
\be
u_w := u(t) .
\ee
In this limit most of the time $\partial_t u(x,t)=0$, and \Eq{EOM}
yields
\be
 F \big(u_w \big) = - m^2 ( w-  u_w )  .
\ee
The geometric construction to find this solution is shown on Fig.~\ref{f:DPM}.
The critical force is   defined as 
\be\label{fc-measured}
f_{\rm c}(m) :=- \overline {F(u_w)} \equiv m^2\overline{ [ w-  u_w ] }. 
\ee
The signs are such that applying the external force $f_{\rm c}$ (counted positive in the driving direction)   overcomes the pinning forces $F(u_w)$.

At a finite driving velocity $v$, averaging \Eq{EOM} yields an additional term $\overline {\dot u} = v$, leading, at least for small velocity $v$, to 
\be\label{fc-v}
f_{\rm c}(m) :=- \overline {F(u_w)}  \equiv m^2\overline{ [ w-  u_w ] }-v. 
\ee
The effective disorder force-force correlator is defined as
\bea\label{16}
\Delta(w-w') &:=&\lim_{v\to 0} \overline {F(u_w) F(u_{w'})}^{\rm c} \nn\\
&=& \lim_{v\to 0}  m^4  \overline{[ w-  u_w ][ w'-  u_{w'} ]}^{\rm c}.
\eea
Note that if we consider an extended system, and $u_w$ is the center of mass position $u_w:= \frac1{L^d} \int_x u_w(x)$, then there is an additional factor of $L^d$ on the r.h.s.\ \cite{MiddletonLeDoussalWiese2006,LeDoussalWiese2006a}.

In a real experiment, it is impossible to measure adiabatically, and instead one measures at a finite velocity, 
\be 
\Delta_v(w-w'):=   m^{4} \overline{[w-u_{w}] [w'-u_{w'}] }^{\,\rm c} .
\label{De-v}
\ee
By definition, 
\be
\Delta(w-w')= \lim_{v\to 0} \Delta_v(w-w').
\ee
While $\Delta(w-w')$ is the second cumulant of the effective action of the field theory  \cite{LeDoussalWiese2006a}, the expectation \eq{De-v} is an observable. 
Perturbation theory allows us to calculate it as
\be\label{Delta-u-theory}
\Delta_v(w) =  \int\limits_0^\infty\!\rmd t \! \int\limits_0^\infty\!\rmd t' \, \Delta(w{-}vt{+}v t' )R(t) R(t'), 
\ee
where $R(t)$ is the response of the center of mass to  an increase in $w$, and $\int_t R(t) = 1$.
(Usually, the response is defined w.r.t.\ an increase $\delta F$ in force. Using $\delta F= m^2 \delta w$, the response w.r.t.\ to a force is normalized as $\int_t R(t) = 1/m^2$.) 

\begin{figure}[h]   
\Fig{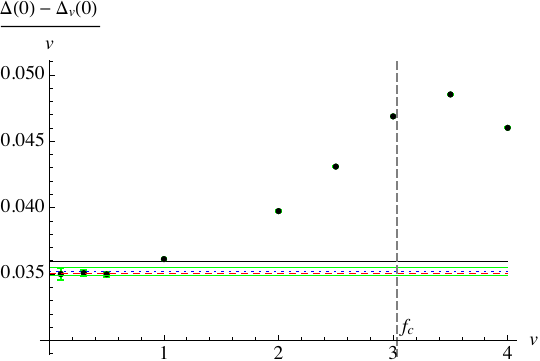}     
\caption{The measured amplitude of the rounding  $v^{-1}[{\Delta_{v=0}(0) - \Delta_{v}(0)}]$ for $m^2 = 0.01$ and $v = 0.1, 0.3, 0.5, 1, 2, 3.5, 4$; $f_{\rm c} \simeq 3.035$. Error bars (green) are included in the fit and represent $99\%$ confidence intervals on the data. We compare several curves. In black is shown the prediction from \Eq{rounding} with $\tau=\tau_m$, and $\Delta'(0^+)$ as given by \Eq{83}. In blue  the prediction from  \Eq{rounding} with $\tau=\left<t \right>$, and $\Delta'(0^+)$ as measured in the simulation, accompanied by  error bars in green. In red  a   weighted fit   to the velocities $v  \leq 0.5$, verifying \Eq{rounding} with a relative precision of $3\times 10^{-3}$. Note the transition at $v=f_c$. (DNS, $\delta u = 0.01$, $\delta t = 10^{-4}$, $N = 10^8$).
}
\label{FigLinearVelScalingMvmax1}
\end{figure}

 \Eq{De-v} implies that  the integral of $\Delta_v(w)$ is independent of $v$.
In general, both $\Delta(w)$ and $R(t)$ may themselves depend on $v$.
We   show below in section \ref{s:The disorder correlator Delta_v(w) at a finite driving velocity} that using the zero-velocity functions on the r.h.s.\ of \Eq{Delta-u-theory} is sufficient at small driving velocities $v$, and the error made is probably $\ca O(v^3)$ or smaller, see Fig.~\ref{FigLinearVelScalingMvmax1}.

An analytic expression for the amplitude of the rounding can be given by expanding \Eq{Delta-u-theory}  at $w=0$ for small $v$,
\bea
&&\Delta_v(0) \nn\\
&&= \int\limits_0^\infty\!\rmd t \! \int\limits_0^\infty\!\rmd t' \, \left[ \Delta(0)+v \Delta'(0^+)|t-t'|+   \ca O(v^2)\right]  R (t) R (t')\nn 
\label{21}
\\
&& = \Delta(0) + v  \tilde\tau  \Delta'(0^+)  +\ca O(v^2),\\
\label{rounding}
&&\tilde \tau :=\int\limits_0^\infty\!\rmd t \! \int\limits_0^\infty\!\rmd t' \,  |t-t'|  R (t) R (t') , 
\label{tilde-tau}
\\ &&\tau := \left< t\right>  \equiv \int_0^\infty \rmd t \, R(t)  t .
\label{tau}
\eea
If $R(t)$ decays exponentially, then $\tilde \tau = \tau$, which should remain a good approximation in most cases.

\begin{figure} [tb]
\Fig{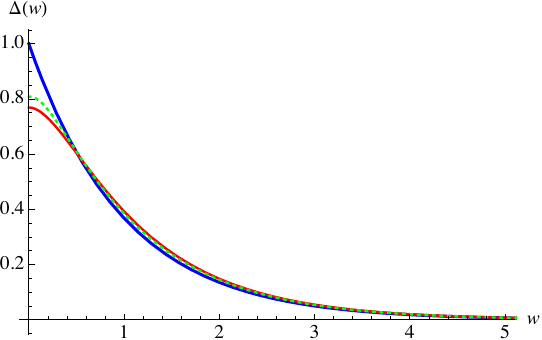}
\caption{$\Delta(w)=\rme^{-w}$ (blue solid), and  the  finite-velocity correlator   \eq{Delta-rounded}  (red) for $\tau v=0.3$, compared to  the boundary-layer approximation \eq{BL} (green dotted).}
\label{f:BL}
\end{figure}

As an illustration for the operation defined in \Eq{Delta-u-theory}, consider $\Delta(w) = \Delta(0) \rme^{-|w|/\xi}$, and 
\be\label{R}
R (t)=\tau^{-1}\rme^{-t/\tau}.
\ee
Then, as plotted on Fig.~\ref{f:BL},
\be\label{Delta-rounded}
\Delta_v(w) = \Delta(0) \, \frac{  
   \rme^{-{|w|/\xi }}-\frac{\tau  v}{\xi}
   \rme^{-{|w|/(\tau 
   v)}}}{1- \big(\frac{\tau  v}{\xi} \big)^2}\ .
\ee
This is a superposition of two exponentials, with the natural scales $\xi$ and $\tau v$. Since
\be
\Delta_v'(0^+)=0,
\ee
the  cusp characteristic for depinning at the origin is rounded. This
  can   be proven   in general  from \Eq{Delta-u-theory}. Note   that $\Delta_v(w)$ is {\em not} analytic, as the expansion contains a term of order $|w|^3$.

Since experiments are performed at a finite driving  velocity, but we are mostly interested in the zero-velocity limit, the question arises of whether 
the folding of \Eq{Delta-u-theory} can be undone.  
There are several possibilities to do this.

\subsubsection{Boundary-layer analysis}
\label{s:Boundary-layer analysis}
As long as  $\tau v\ll \xi$, the second term of \Eq{Delta-rounded} decays much faster than the first, allowing us to perform a {\em boundary-layer analysis}.
This term was coined in the context of turbulence, where a turbulent bulk behavior has to be connected to a laminar boundary layer. 
There is a large mathematics and physics literature on the subject. Relevant keywords are  {\em boundary layer} (physics literature) or {\em singular perturbation theory} (mathematics literature); a few references to start with are \cite{Wasow1965,Bogolyubov2011,Smith1985,HairerWanner1996}.
Let us proceed   by noting that \Eq{Delta-u-theory} can be approximated by   the   boundary-layer ansatz 
\bea\label{BL}
\Delta_v(w) \simeq \ca A_v\, \Delta \Big(\sqrt{w^2+\delta _w^2}\Big),\quad \\
\delta  _w =\tau v,\quad \tau := \left< t\right>  =\int_0^\infty \rmd t \, R(t)  t \ ,\\
\ca A_v = \frac{\int_0^\infty\rmd w\, \Delta(w)}{\int_0^\infty\rmd w\, \Delta(\sqrt{w^2+ \delta_w ^2})}.\quad
\eea
The   amplitude $\ca A_v$ ensures   normalization. The quality of this approximation can be seen on Fig.~\ref{f:BL}: it works well for $v$ small, but deteriorates for larger $v$. 

We can 
use the boundary-layer formula \eq{BL} to plot the measured $\Delta_v(w)$ against $\tilde w := \sqrt{w^2+ \delta _w ^2}$; and then   find the best  $\delta_ w$ which removes the curvature of $\Delta_v(w)$. It yields $\delta _w$, and by extrapolation to $w=0$ the full $\Delta(w)$. This idea is tested below in section \ref{s:reconstruction-simul}.

\subsubsection{Estimate of time scale}
If in an experiment the response function is unavailable, using the boundary-layer ansatz \eq{BL}, its characteristic time scale $\tau$ can  be reconstructed approximatively from $\Delta_v(w)$ as 
\be\label{vtau-estimate}
\delta_w = \tau v \simeq \frac{\lim_{w\to 0 }\Delta'(w)}{\Delta_v''(0)}.
\ee
In the numerator is written  $\lim _{w\to 0}\Delta'(w)$, which  is obtained by extrapolating $\Delta_v'(w)$ from outside the boundary layer, i.e $w\ge \delta  _w = \tau v$, to $w=0$.

\subsubsection{Differential equation}
\label{s:Differential equation}
First note that the response function $R(t)$ defined in \Eq{R} satisfies the differential equation 
\be\label{R-diff}
(\tau \partial_t +1)  R(t) = \delta(t).
\ee
Second,   rewrite  \Eq{Delta-u-theory} as
\bea
&&\Delta_v\big(v(t-t')\big) \nn\\
&&=  \int\limits_{-\infty}^t \rmd t_1  \int\limits_{-\infty }^{t' }\rmd t_2 \, \Delta\big(v(t_1{-}t_2)\big)R (t-t_1) R (t-t_2). ~~~
\eea
Applying the differential operator of \Eq{R-diff} both  to $t$ and $t'$ yields
\be
(\tau \partial_t +1) (\tau \partial_{t'} +1) \Delta_v\big(v(t-t')\big) = \Delta\big(v(t-t')\big).
\ee
In terms of the variable $w$, this relation can be simplified to 
\be\label{242}
\Delta(w) = \left( 1 - (\tau v)^2 \partial_w^2 \right) \Delta_v(w).
\ee
While \Eq{242} is more precise, and reconstructs $\Delta_{v=0}(w)$ down to $w=0$, the boundary-layer analysis may be more robust for noisy data.  We will examine these procedures in section \ref{s:reconstruction-simul}.

\subsubsection{Other auto-correlation functions}
\label{s:Other auto-correlation functions}
\Eq{EOM}   allows us to consider three different observables,
\bea
\label{uw}
u_w &=& u(t),\\
\label{udotw}
\dot u_w &=& \dot u(t) ,\\
\label{Fw}
F_w &=&F(u_w)= F\big(u(t)\big) .
\eea
What is measured in magnetic domain-wall experiments   is the induced current, proportional to $\dot u_w$ \cite{UrbachMadisonMarkert1995,DurinZapperi2000,KimChoeShin2003,DurinZapperi2006b}; in contact-line depinning where the interface is filmed, this is $u_w$ \cite{LeDoussalWieseMoulinetRolley2009};   when stretching an elastic material as a knit, this is  the {\em external} force  exerted on the knit, itself proportional to  $u_w-w$ \cite{PoinclouxAdda-BediaLechenault2018}. We do not know of any system where one can measure  solely the force $F_w$ of the disorder.

\Eqs{uw}-\eq{Fw}
define three auto-correlation functions
\bea
\label{Delta-v}
\Delta_v(w-w')&:=& m^4\overline{ [u_w-w][ u_{w'}-w')]}^{\rm c}, \qquad\\ 
\label{Delta-udot}
\Delta _{\dot u}(w-w')&:=&  \overline{ [\dot u_w-v][\dot u_{w'}-v]}^{\rm c}, \\ 
\label{Delta-F}
\Delta _F(w-w')&:=& \overline{ F_w F_{w'}}^{\rm c} .
\eea
We will show in section \ref{s:3 Deltas} that they satisfy
\bea
\Delta_F(w)  &=&  \Delta_v(w)+  \Delta_{\dot{u}}(w)   \label{UFUdotRelation}, \\
\Delta_{\dot u}(w) &=& - \frac{v^{2}}{m^{4}} \partial_w^2  \Delta_v(w).
\eea

\section{Review of known results for ABBM and DPM}
Key features for the ABBM model and the DPM are given on table \ref{t:compare}.

\begin{table*}[t]
\centering 				
\renewcommand{\arraystretch}{1.7}					 
\begin{tabular}{|c | c | } 								
\hline 									
\quad	\qquad\qquad	\qquad\qquad ABBM model	\qquad	\quad\qquad\qquad	\qquad	& \qquad	\qquad Discretised Particle model 	(DPM) \qquad\qquad	\\ 

\hline
\multicolumn{2}{|c|}{characteristic scale of effective (renormalized) force correlator} \\
\hline
1 & $\rho_{m} =  \Big(  m^2
   \sqrt{2\log
    ( m^{-2} )} \Big) ^{-1} $ \rule[-1.7ex]{0mm}{5ex}\\
\hline
\multicolumn{2}{|c|}{effective (renormalized) disorder $\Delta(w)$ }	 \\
\hline
$\Delta(0)-\Delta(w) =  \sigma |w| 	$, $\sigma=1$   &	\qquad $ \Delta(w) =  m^4 \rho_m^2   \left[ \text{Li}_2(1 - \rme^{|w|/\rho_m}) + \frac{w^2}{2\rho_m^2}+ \frac{\pi^2}{6} \right]$  \rule[-1.9ex]{0mm}{5ex} \qquad	\\
\hline 		
\multicolumn{2}{|c|}{ critical force }  \\
\hline
$f_{\rm c}(m^2 ) = F(u=0)	$	&  	$ f_{\rm c}(m^2) = \sqrt{2\log (m^{-2}) }$ 	\\ 	
\hline 											
\multicolumn{2}{|c|}{ response function $R (t) =  {\tau}^{-1} \rme^{-t/\tau}$	} \\
\hline
$\tau^{-1} = m^2$    	&  $\tau^{-1}= \tau_m^{-1} :=  2m^2\log( m^{-2} ) $ 	\\ 
\hline 											
\multicolumn{2}{|c|}{avalanche-size distribution for infinitesimal kick $P(S) \sim S^{-\tau} e^{-S/S_m}$ }	 \\
\hline
$ \tau = 3/2 $, $S_m = m^{-4}$ 	&    	$ \tau = 0 $,  $S_m = \rho_m  $	\\
\hline 											
\multicolumn{2}{|c|}{avalanche-duration distribution for infinitesimal kick $P(T) \sim T^{-\alpha} e^{-T/T_m}$}	 \\
\hline
$\alpha=2$, $T_m =\tau=m^{-2}$	  &         $\alpha=0	$, $T_m= \tau_m = [ 2m^2\log( m^{-2} )]^{-1}$  \\
 [1ex] 	
\hline 	
\multicolumn{2}{|c|}{roughness exponent $\zeta$, defined by $u\sim m^{-\zeta}$}	 \\
\hline
$\zeta=4$  &         $\zeta=2^{-}$ \quad(2 \mbox{reduced by logarithmic corrections})\\
 [1ex] 	
\hline 											
\end{tabular}
\caption{Comparison of the ABBM model with the DPM.}
\label{t:compare} 										
\end{table*}

\subsection{ABBM model}
The response function is unchanged from the free theory
\be\label{R-ABBM}
R(t) = m^2 \rme^{-m^2 t} \Theta(t).
\ee
The  velocity distribution $P_t(\dot{u})$ reads
\cite{AlessandroBeatriceBertottiMontorsi1990,Bertotti1998,Colaiori2008,DobrinevskiLeDoussalWiese2011b}
\be
P_t(\dot{u}) = m^2 \left( m^2 {\dot{u}} \right)^{v m^2- 1}  \frac{\rme^{-m^2 \dot{u}}}{\Gamma ( v m^2)}.
\label{ABBMformulaV}
\ee
By construction it is normalized,   its first moment is   $\left< \dot u \right> = v$, and 
\be
v_m ^{\rm ABBM}:= \frac{\langle \dot u^2\rangle}{\langle\dot u \rangle} = m^{-2} + v.
\ee
The avalanche-size distribution $P_{\delta w}^S(S)$, given a kick $\delta w$, reads \cite{DobrinevskiLeDoussalWiese2011b}
\be\label{PwS(S)=ABBM}
  {P_{\delta w}^S(S) = m^2 \delta w  \frac{ e^{-\frac{m^4 (S-\delta w)^2}{4
      S}}}{2 \sqrt{\pi    } S^{3/2}} }. 
\ee
The avalanche-size exponent is $\tau=3/2$. 
The first moments are 
\be
\left< 1 \right> = 1, \quad \left< S \right> =\delta w, \quad S_m^{\rm ABBM} := \frac{ \left< S^2 \right>}{2  \left< S \right>} =  m^{-4}+ \frac{\delta w} 2. 
\ee
The   duration distribution given a kick of size $\delta w$ is    
\be\label{P-duration1}
P_{\delta w}^T(T) 
=  \exp\!\left(-\frac{\delta w m^4}{\rme^{Tm^2}-1} \right)  \frac{\delta w m^6}{[2\sinh(Tm^2/2)]^2}.
\ee
This distribution is normalized. The first   moments   
 to leading order in $\delta  w$ are 
\bea\label{T-ABBM}
\left< T\right>    &=& m^2   \Big[ 1 -\gamma_{{\rm E}} - \ln (m ^4   \delta w)   \Big]+ ... \\
   \left< T^{2}\right> &=& \frac{\pi^{2}}{3} \delta w   +... \\
     \left< T^{3}\right> &=&{6 \zeta(3)} \frac{\delta w}{m^{2}}+ ...\\  
     \left< T^{4}\right> &=&\frac{4 \pi^4}{15} \frac{\delta w}{m^{4}}+ ... \\
 T_{m}^{\rm ABBM}&:=&\frac{\left< T^{3}\right>}{3\left< T^{2}\right>}   =\frac{6  \zeta (3)}{\pi^2 m^{2}}+ ...=\frac{0.730763}{m^{2}}+...~~~~
\eea

\subsection{The discretized particle model (DPM)}
The discretized particle model (DPM) was introduced in Ref.~\cite{LeDoussalWiese2008a}. There the reader  finds a thorough discussion of its quasi-static properties, encompassing all three main universality classes of extreme-value statistics: Gumbel, Weibull, and Fr\'echet.

\subsubsection{Static quantities}
As we assume forces to be distributed according to
\be\label{PF}
{  P}_F(F)= \frac{\rme^{-F^2/2}}{\sqrt{2\pi}}  ,
\ee
this is the Gumbel class of Ref.~\cite{LeDoussalWiese2008a} with $A=1/2$,  $\gamma=2$,  $\beta(x)=x^2/2$, and $\beta^{-1}(x) =\sqrt{2 x}$.
The name of this class stems from the fact that the blocking    forces are distributed according to a {\em Gumbel} distribution (\cite{LeDoussalWiese2008a} Eq.~(29), first line)
\bea
P_{\rm G}(a) &=& \rme^{-a}\Theta(a) ,\label{PG}\\
f&=& \sqrt 2 \sqrt{ \ln (m^{-2}) -  \ln(a)} \label{f-a}\\
 &=& f_{\rm c}^0  - \ln (a)\, m^2 \rho_m + ... 
\eea
The constant $f_{\rm c}^0$, the scale $\rho_m$, and the exponent $\zeta$ are
\bea
f_{\rm c}^0  &=&   \sqrt{2 \ln(m^{-2})} + \ca O(m^{-1}),   \nn \\
 \rho_{m} &=& \frac{1}{  m^2
   \sqrt{2\log
    ( m^{-2} )}
   }, 
\label{rho-m}   
   \\
   \zeta &=& 2^{-}.
\eea
By $\zeta=2$ we  mean $\rho_m \sim m^{-\zeta}$ with $\zeta=2$.  The growth of \Eq{rho-m} with $1/m$ is slightly slower, reduced by the logarithm in \Eq{rho-m}, and  denoted $\zeta = 2^{-}$.

The effective disorder force-force correlator reads  
\bea
 \Delta(w) &=& m^4 \rho_m^2 \tilde \Delta (w/\rho_m) ,  \label{formdelta}\\
\tilde \Delta (w) &=& \frac{w^2}2+\mbox{Li}_2(1-\rme^{w})+\frac{\pi^2}6.
\eea
The avalanche-size distribution for an infinitesimal kick was obtained  in Ref.~\cite{LeDoussalWiese2008a}, where it was shown to be $P(S) \sim \rme^{-S/\rho_m}$. For a   kick of size $\delta w$, this can be generalized to
\be\label{PofS-discrete}
P_{\delta w}^{\rm DPM}(S) = 4 { \delta w}   \sinh\!\left(\frac1{2\rho_m}\right)^{\!2}  \rme^{- S/\rho_m}\simeq \frac{\delta w}{\rho_m^2} \rme^{- S/\rho_m} \ .
\ee
Note that for the DPM, the avalanche size $S$ is discrete. The normalization is constructed s.t.\ the first moment of this discrete measure is
$
\left< S \right> = \delta w.
$
This yields for the characteristic  scale of avalanches
\be
S_m = \frac{\left< S^2\right> }{2 \left< S \right>} = \frac12 \coth\!\left(\frac1{2\rho_m}\right)  =  \rho_m + \ca O(1/\rho_m) \ .
\ee

\subsubsection{Dynamic quantities} The DPM defined in Ref.~\cite{LeDoussalWiese2008a}   advances instantaneously. The easiest way to endow it with a realistic dynamics is to consider the Langevin equation \eq{EOM}.
If the disorder is needle-like as on the right of Fig.~\ref{f:DPM} (the original construction of \cite{LeDoussalWiese2008a}), then either the particle is at rest blocked by a needle, or it moves, and the only force acting on it  comes from the spring. Neglecting that the spring gets shorter during the movement, the response-function is   given by 
$
R(t) \sim  P(S/v)
$, where $  v =f_{\rm c}$, resulting   in
\be\label{R(t)DPM}
R^{\rm DPM}(t) = \tau_m^{-1} \rme^{-t/\tau_m}, \quad \tau_m =  \frac{\rho_m}{f_{\rm c}}= \frac1{2 m^2 \ln(m^{-2})} .
\ee 
Stated differently, the velocity distribution is 
\be\label{P-udot-DPM}
P^{\rm DPM}(\dot u) = \delta(\dot u- f_{\rm c}) .
\ee
As a consequence, the  distribution of durations $T$, given a kick $\delta w$, can be obtained from the avalanche-size distribution as 
\be\label{P(T)DPM}
 P_{\delta w}^{\rm DPM}(T) \simeq \frac{ \delta w} {\rho_m \tau_m} \rme^{-T/\tau_m}\ . 
\ee

\section{Numerical results}
\subsection{Critical Force}
\begin{figure}[t]
\Fig{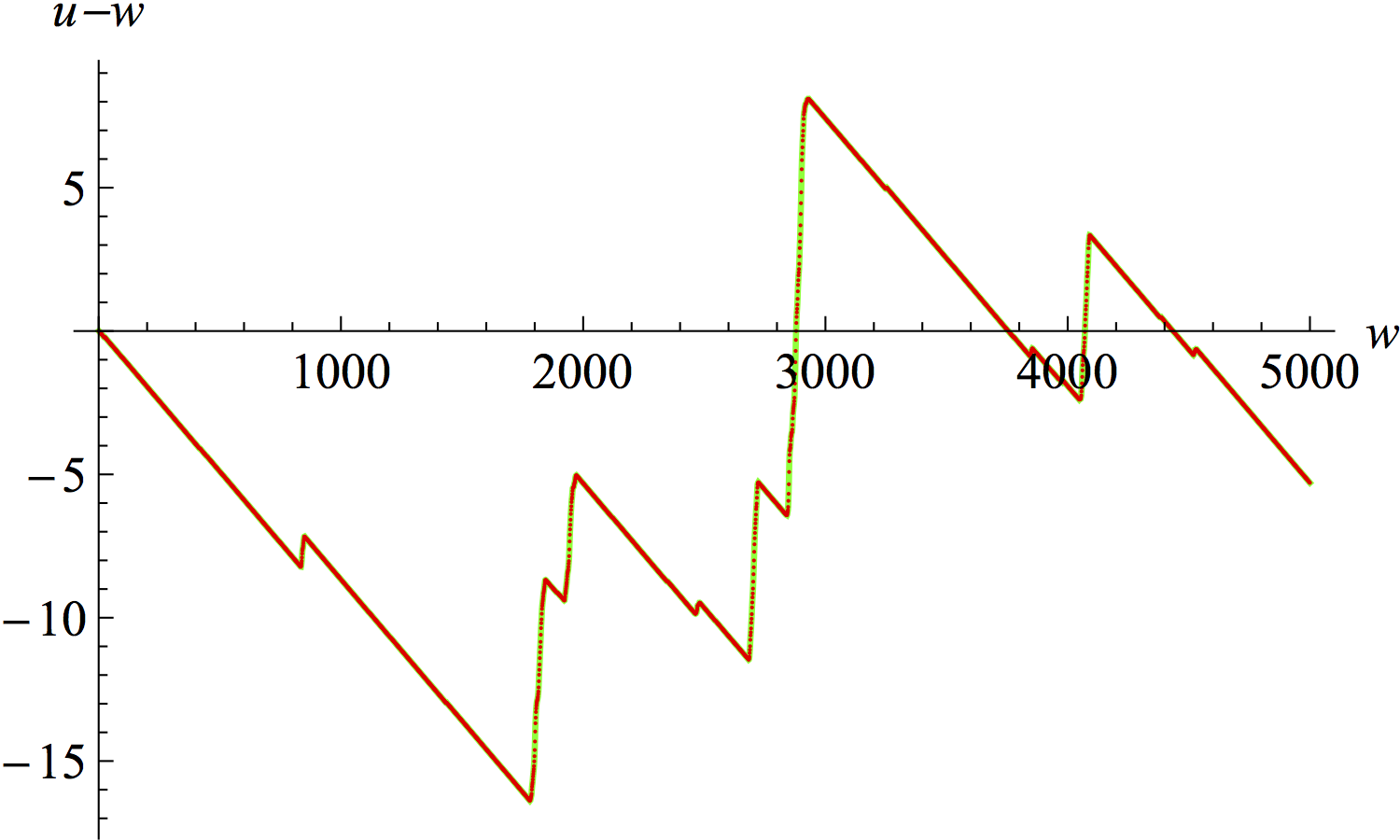}
\caption{$u_w-w$ for $\delta t = 10^{-4}$, 
$ v = 0.1$, 
$  m^2 = 0.1$, INS. See section \ref{s:Numerical implementations} for implementation details. Red points are data points equally spaced in time, the green line between them is an interpolation.}
\label{f:u(w)}
\end{figure}
\begin{figure}[t]
\Fig{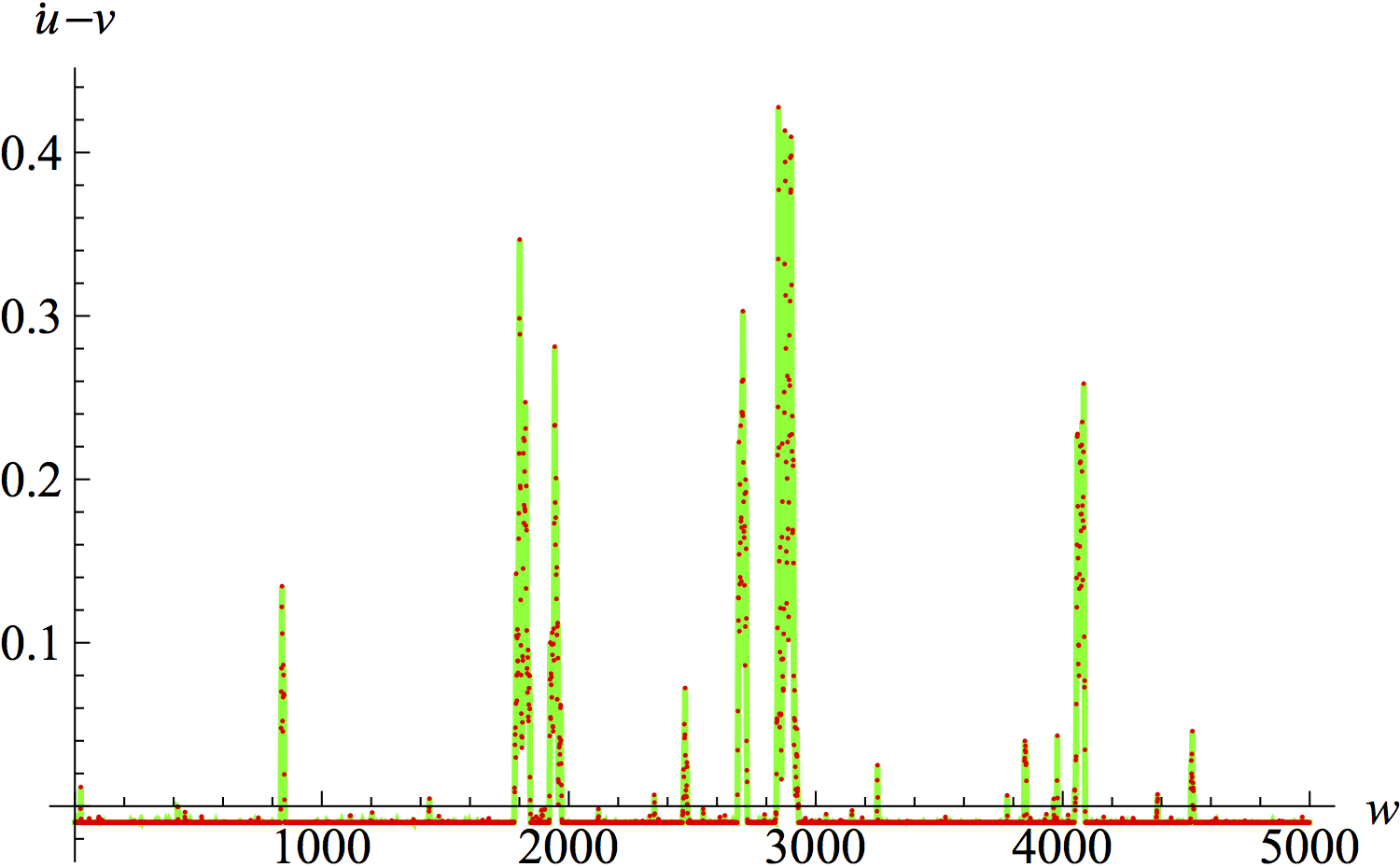}
\caption{$\dot u(w)-v$ with the same parameters as in Fig.~\ref{f:u(w)}.}
\label{f:udot(w)}
\end{figure}
\begin{figure}[t]
\Fig{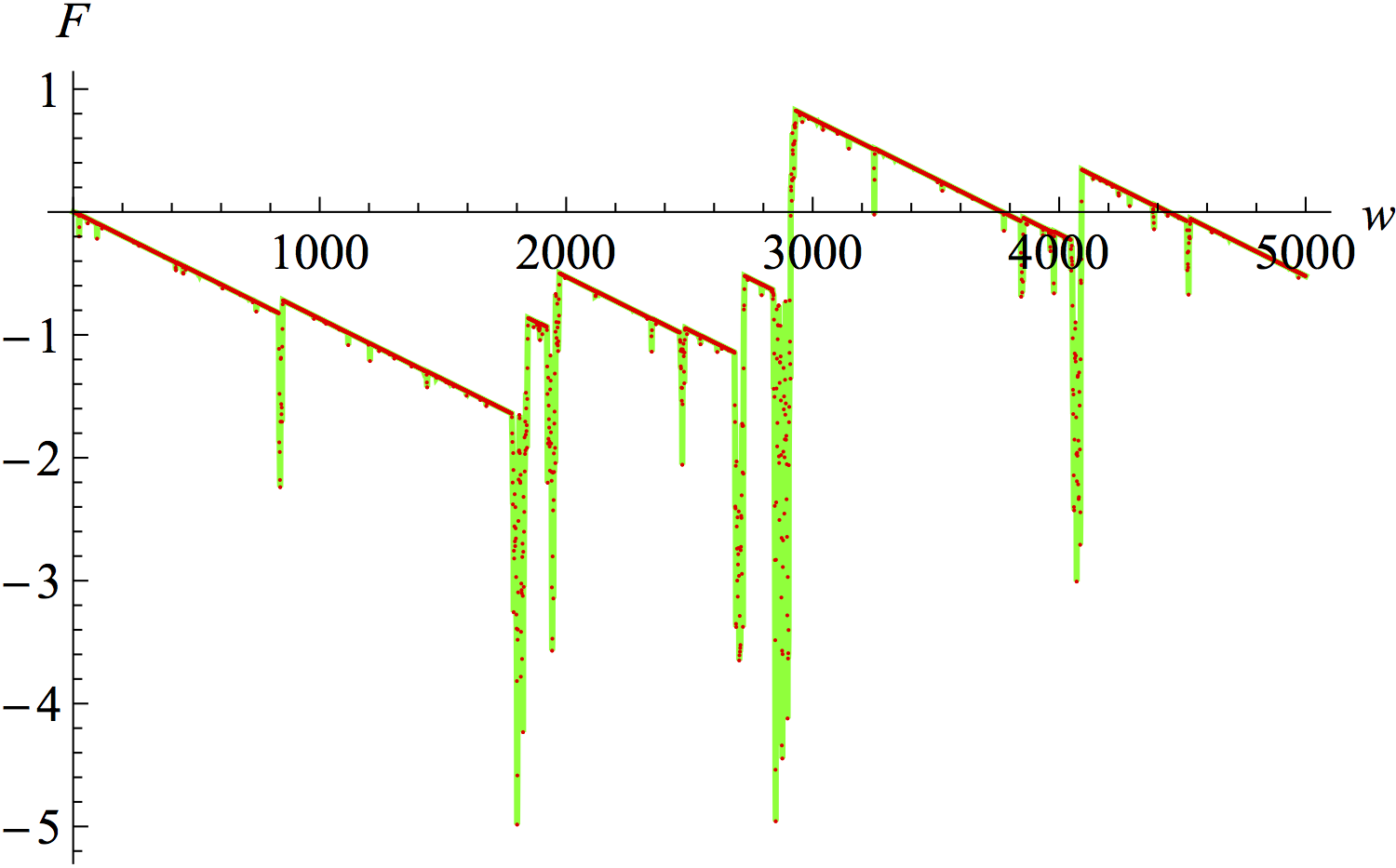}
\caption{$F(w)$ with the same parameters as in Fig.~\ref{f:u(w)}.}
\label{f:F(w)}
\end{figure}
First one integrates the
  equation of motion \eq{EOM}, with forces as given by \Eq{Ornstein-Uhlenbeck}. This gives $u(t)$, or $u_w$ as a function of the external point $w=v t$,  defined s.t.\ $u_w=u(t)$. Example trajectories for $u_w-w$, $\dot u_w-v$, and $F_w$ are plotted in Figs.~\ref{f:u(w)}-\ref{f:F(w)}. 

It is important to note that the position $u_w$ and force ${ F}_w$ are related, since \Eq{EOM} yields after averaging over time 
\be
\overline{\partial_{t } u(t)} =  v = m^{2} \overline{[w-u(t)]} +  \overline {F(t)}.
\ee
Note that the overline, defined as an average over disorder realizations, can be performed as an average over time when driving the system at a finite velocity $v>0$. 
\begin{figure}[t]   
\includegraphics[width=\columnwidth]{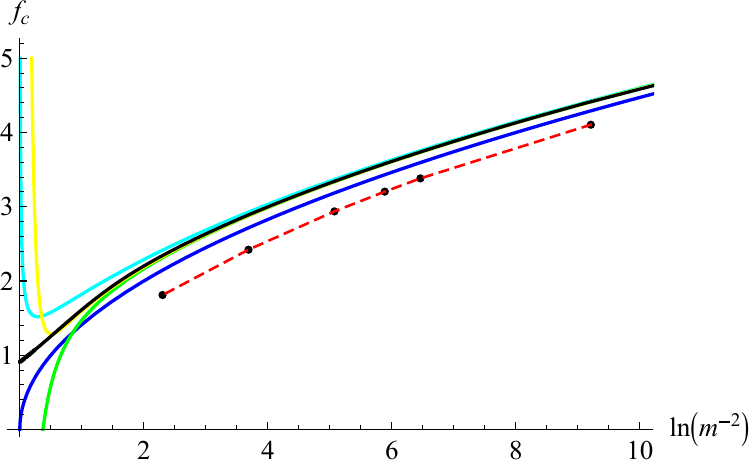}
\caption{The critical force $f_{c}$ as a function of $m$. Thick black: the results of the integral \eq{FcTheoryFull}. The small-$m$ expansion is in blue, cyan, green, and yellow. The data points are from our numerical simulation.   (DNS, $\delta u=10^{-2}$,  $\delta t=10^{-4}$, $N=3\times 10^6$.)}
\label{figFcrit}
\end{figure}
For the discretised particle model, the critical force was computed analytically in Ref.~\cite{LeDoussalWiese2008a}. It is given by 
\Eqs{PG}-\eq{f-a}
 \begin{equation}
f_{\rm c}(m) := \left<f\right>_{\rm G} = \sqrt{2} \int_{0}^{m^{-2}}  \sqrt{  \ln(m^{-2}) - \ln(a) } \;\rme^{-a}\,\rmd a \ .
\label{FcTheoryFull}
\end{equation} 
Expanding for small $m$, we find 
\bea
\nn
f_{\rm c} (m) & = & f_{\rm c}^0 + \frac{\gamma_{\rm E}}{f_{\rm c}^0}  - \frac{\frac{\gamma_{\rm E}^2}{2} + \frac{\pi^2}{12}}{(f_{\rm c}^0)^3}  \label{Fexpansion1}\\ 
\ &&+   \frac{2 \gamma_{\rm E}^3 + \gamma_{\rm E} \pi^2 - 2 \psi''(1)}{4(f_{\rm c}^0)^5}  + ...  \\ 
 f_{\rm c}^0  &=& \sqrt{2 \log(m^{-2})} .
\eea
In Fig.~\ref{figFcrit} we compare the measured critical force for the Ornstein-Uhlenbeck model, 
$f_{\rm c}^{\rm OU} \equiv m^2 \overline{[w -u(w)]} - v$ 
 to the critical force \eqref{Fexpansion1} predicted by the particle model. 
 We find that they agree, up to a constant
\begin{equation}
 f_{\rm c}^{\rm OU} \simeq f_{\rm c}(m) - 0.35  \label{FcShift} .
 \end{equation}
This constant  is not surprising, as the microscopic disorder of the Ornstein-Uhlenbeck process is  different from the DPM, and the critical force is not universal. We note that the reported value $0.35$ is almost the correlation $\rme^{-1} \approx 0.368$ of the Ornstein-Uhlenbeck process at distance $u=1$.

\subsection{Velocity distribution}
 
\begin{figure}[t]
\Fig{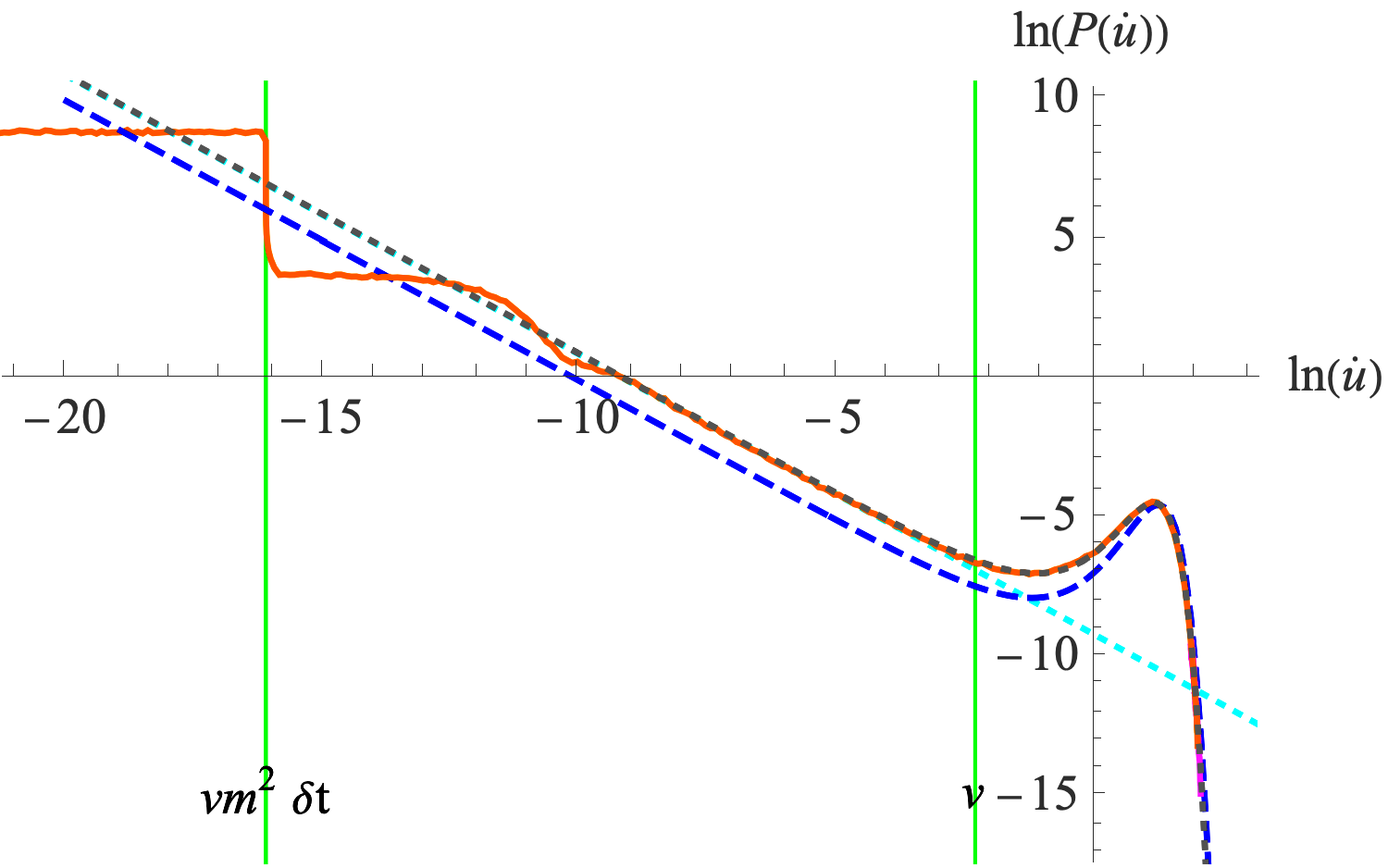}
\caption{$P(\dot u)$ for $m^{2}=10^{{-3}}$ (orange, solid line). The blue dashed curve is given by \Eq{P-of-udot}, the gray dotted one by the same formula, replacing $m^{-2}\to 0.39 m^{-2}$, yielding a perfect fit. Cyan dotted is  the result  for ABBM, \Eq{ABBMformulaV}. (INS, $\delta t=10^{-3}$, $v=0.1$, $N=5 \times 10^{8}$).}
\label{f:P(dotu)-imp}
\end{figure}

In the discretized-force model, and supposing that the forces remain constant between integers, the velocity distribution  between site $u$ and $u+1$ is given by 
\be
P_{u}(\dot u)= \left< P_{ F}(f -\dot u) \right> _{\rm G } .
\ee
Here $P_F(f)$ is the initial force distribution \eq{PF}, and the average is over the Gumbel distribution   defined by \Eqs{PG}-\eq{f-a}.
As we are interested in the velocity distribution sampled equally   in time and not in space, we still have to multiply with $v/\dot u$, resulting in
\bea\label{P-of-udot}
P_t(\dot u) &=& \frac{v}{\dot u} P_{u}(\dot u)   \\
&=& \frac v{\dot u}\frac 1{\sqrt{2\pi}} \int_0^{m^{-2}}\rme^{- \left(\frac{\dot u}{\sqrt 2} - \sqrt{\ln(m^{-2}) - \ln(a)} \right)^2 } \rme^{-a}\, \rmd a. \nn
\eea
This formula, evaluated numerically, is compared to simulations for $m^2=10^{-3}$  on Fig.~\ref{f:P(dotu)-imp}. While the  
tail is correctly predicted, the amplitude for small $\dot u$ is underestimated. This is related to the underestimation of $f_{\rm c}$ reported in \Eq{FcShift}. Indeed, we find a perfect fit for $m^2=10^{-3}$ under the replacement $m^{-2} \to 0.39 m^{-2} $, see Fig.~\ref{f:P(dotu)-imp}.

\subsection{The response function}
\label{s:The response function}
\subsubsection{Measurement prescription}
\label{s:{Measurement prescription}}
The response function  plays a key role as it enters into the rounding of the cusp, and we need to measure it to verify the prediction in \Eq{R(t)DPM}.
To this aim, we let the system relax to $\dot{u} = 0$ and then   kick it at time $t= t_{\text{kick}}$, moving the center of the well from $w$ to $w + \delta w$. The non-linear response function is then given by
\begin{equation}
R  (t|{\delta w}) :=  \frac{\langle \dot{u}(t + t_{\rm kick}) \rangle  }{\delta w} .
 \label{Rtdef-nonlinear}
\end{equation}
The linear response is the limit of a small kick $\delta w$, 
\begin{equation}
R  (t) := \lim_{\delta w  \to 0} R  (t|{\delta w}).
 \label{Rtdef}
\end{equation}
In practice, $\delta w$ can be chosen finite; we give suitable values on Figs.\ \ref{figresponseDFmodel} and \ref{f:response-OU}.
The response function is normalized, and its first moment defines a characteristic time scale $ \tau $, 
\begin{align}
\int_0^\infty R (t) \, \rmd t = 1, \qquad \tau:= \langle t \rangle = \int_0^\infty t \text{ } R (t)  \,\rmd t . \label{RtConditions}
\end{align}
Comparing the measured time scale  $\tau:=\langle t \rangle $ to the predicted one $\tau_m$ from \Eq{R(t)DPM} is an important check of the theory. 

\subsubsection{Response function for DPM}
\begin{figure}[t]  
\setlength{\unitlength}{1mm} 
\Fig{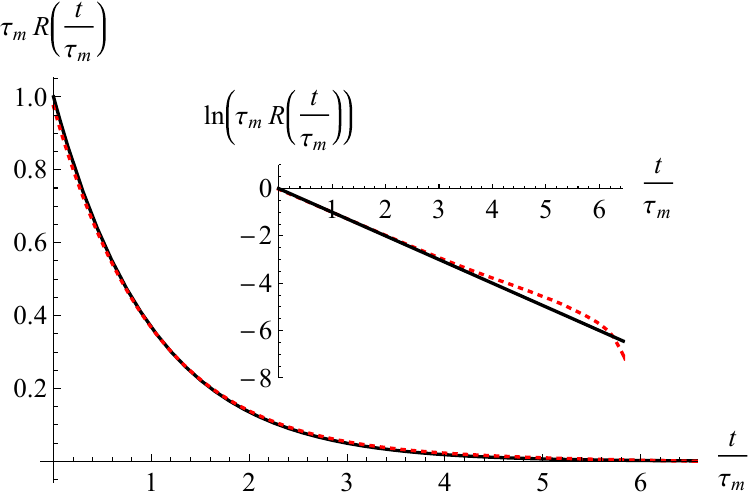}
\caption{The response function for the discretized force model (needle disorder) for $m^2 = 10^{-2}$ (blue).
The red dashed curve is an exponential function  with time scale  $\tau_m $ as given by \Eq{R(t)DPM}. Inset: ibid for $\ln R(t)$.   (DNS, $\delta t=10^{-4}$,    $\delta w=0.1$.)}
\label{figresponseDFmodel}
\end{figure}
We start with the response function for the discretised force model of Ref.~\cite{LeDoussalWiese2008a}. On Fig.~\ref{figresponseDFmodel} we show a numerical verification for the analytic prediction in \Eq{R(t)DPM}. Already for a rather large mass of $m^2=10^{-2}$ the agreement between theory and simulation is rather good.

\subsubsection{Response function for Ornstein-Uhlenbeck forces}
\begin{figure}[t]
\Fig{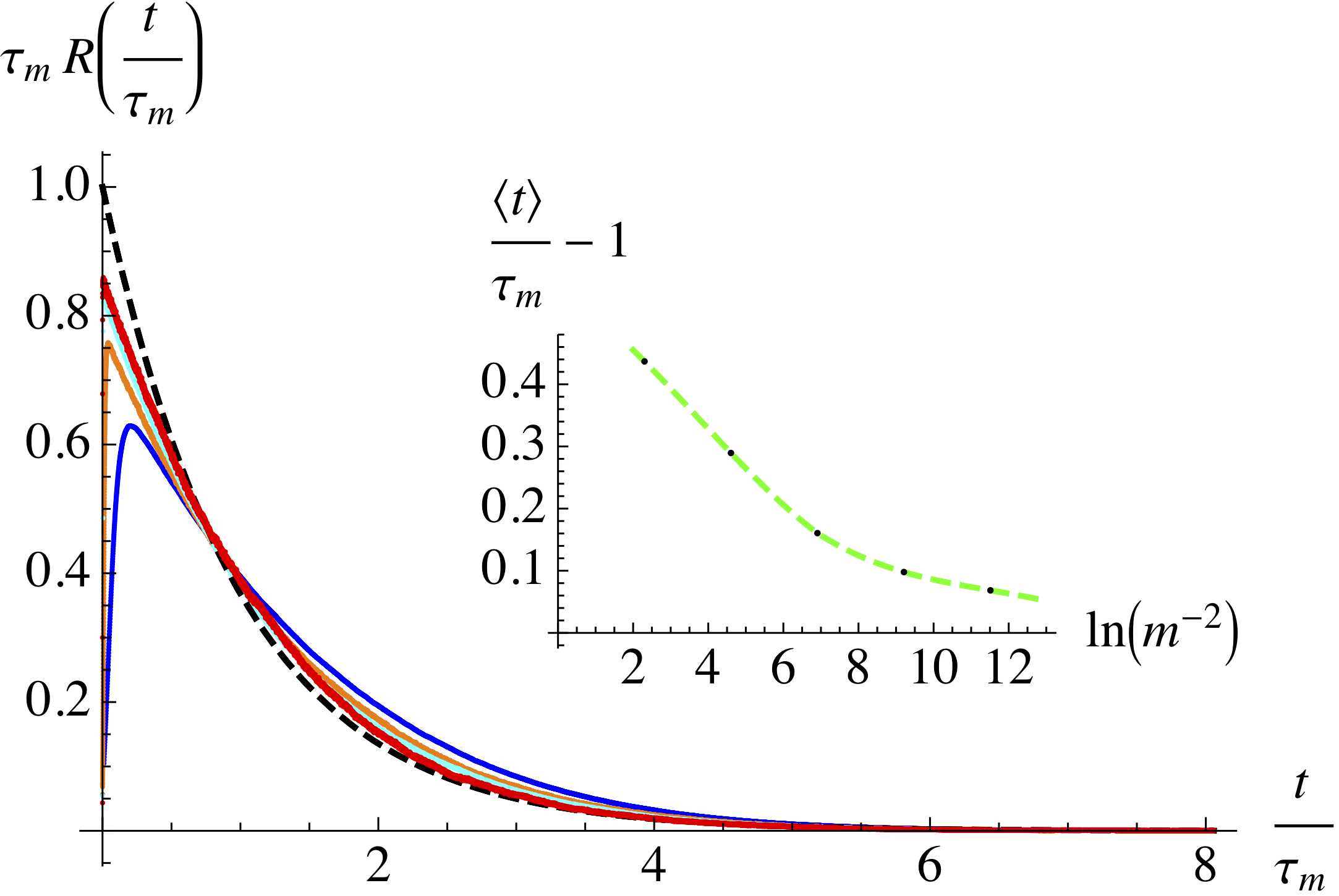}
\caption{The response function for the Ornstein-Uhlenbeck process: analytical result in black dashed. Numerical results from bottom to top are for $m^2=10^{-2}$ (blue), $m^2=10^{-3}$ (orange), $m^2=10^{-4}$ (cyan), and $m^2=10^{-5}$ (red).  DNS,  $\delta t=10^{-4}$, $N=10^8$. The two larger masses have  $\delta w=0.1$ and $\delta u=10^{-6}$, the two smaller ones $\delta w=1$, and $\delta u = 10^{-4}$. Inset: The measured timescale $\langle t \rangle $ (dots) compared to the prediction for $\tau_m$ in  \Eq{R(t)DPM}.}
\label{f:response-OU}
\end{figure}

\begin{figure}[b]   
\includegraphics[width=\columnwidth]{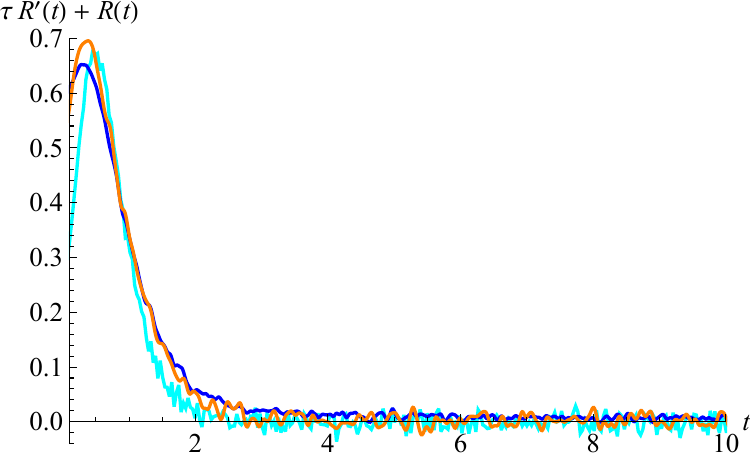}
\caption{The function $\delta_{\rm BL}(t) := (\tau \partial_t +1)  R (t)$ for  the measured  response function with $m^2 = 10^{-2}$ (blue),     $m^2 = 10^{-3}$ (orange), $m^2 = 10^{-4}$ (cyan). Parameters as in Fig.~\ref{f:response-OU}, except $N=2\times 10^8$ for $m^2=10^{-4}$.}
\label{FigCheckUnfolding}
\end{figure}

On Fig.~\ref{f:response-OU}
we show the response function for   Ornstein-Uhlenbeck forces. While for the DPM which has needle-like disorder, the response function on Fig.~\ref{figresponseDFmodel} had already converged  to the asymptotic behavior of \Eq{R(t)DPM}  for $m^2=10^{-2}$, the  convergence for   Ornstein-Uhlenbeck forces is much slower, and one has to go to $m^2=10^{-5}$ to reach a similar agreement, albeit   with a noticeable difference in the time scale $\tau=\left< t \right>$. The  rather slow convergence \cite{terBurgPhD} is shown in the inset of Fig.~\ref{f:response-OU}. A second observation is that the response function $R(t)$ for $t\to 0$ starts at  $1/\tau_{\rm ABBM} = m^2$. This is a consequence of the continuity of $F(u)$: Since in the beginning $m^2[w-u(t)]+F\big(u(t)\big)=0$, the response function is that of the free theory, itself equivalent to that of the ABBM model,  
\be
R(t)\approx R_{\rm free}(t)\equiv  R_{\rm ABBM}(t)  = m^2 \rme^{-m^2 t}\Theta(t).
\ee 
The position $u(t)$ then increases, and since pinning occurs at maxima of $F(u)$, most likely $F(u)$ decreases, leading to an increase in $\dot u$ as compared to the free theory. When $u(t) -u(t_{\rm kick}) \simeq 1$, we expect the forces $F(u)$ to be decorrelated from its initial value. This estimates the boundary layer in $u$ as $\delta u \approx 1$. It is non-trivial to predict the boundary layer in $t$, i.e.\ the region in which $R(t)$ rises, before  decaying approximately as an exponential.  
After some experimentation, we found that this can   be extracted by plotting the combination
\be\label{R-diff-BL}
\delta_{\rm BL}(t) := (\tau \partial_t +1)  R (t) 
\ee
appearing in \Eq{R-diff}, with $\tau$ defined in \Eq{RtConditions}.
By construction
\be
\int_0^\infty \rmd t\, R(t) = \int_0^\infty \rmd t \, \delta_{\rm BL}(t)  =1.
\ee
If $R(t)$ is   exponentially decaying,   $\delta_{\rm BL}(t)$ equals $\delta(t)$. Deviations  lead to  a smeared-out $\delta$-function. On Fig.~\ref{FigCheckUnfolding} we show $\delta_{\rm BL}(t) $ for several masses and time scales ranging from $\tau \approx 14$ to  $\tau \approx 596$. 
Despite the enormous range of time scales, the resulting $\delta_{\rm BL}(t)$ is almost independent of $\tau$, and decays to zero on a range of $\delta t \approx 2$. We believe that this function could be extracted from extreme-value statistics, by considering the statistics of   a Brownian motion close to one of its records. It is e.g.\ known that the fractal dimension of the record set, i.e.\ the position when the movement of the particle stops again, is $1/2$ \cite{MortersPeres2010,BenigniCoscoShapiraWiese2017}.

\subsubsection{The response function at a finite driving velocity}

\begin{figure}
\fig{0.49}{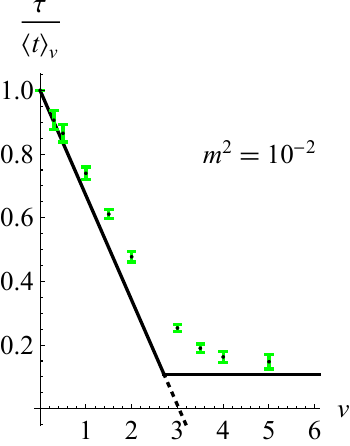}~~\fig{0.49}{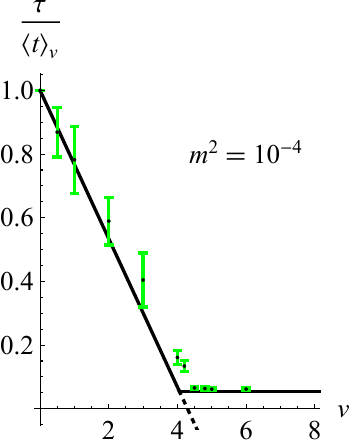}
\caption{Inverse of the measured time scale $\left< t\right>_{v}:= \int_0^\infty \rmd t\, t R_v(t)$, as a function of $v$. The data points are given with $99\%$ confidence intervals (green). Superimposed are the two analytical curves \eq{tau(v)}, valid for $v \le f_{\rm c}$, and \eq{tv}, valid for $v\ge f_{\rm c}$.  (DNS, $\delta t = 10^{-4}$, $\delta u = 10^{-6}$ for $m^2=0.01$, and $\delta u = 10^{-4}$ for $m^2=10^{-4}$.)}
\label{figTimescale}
\end{figure}
\begin{figure}[b]   
\includegraphics[width=\columnwidth]{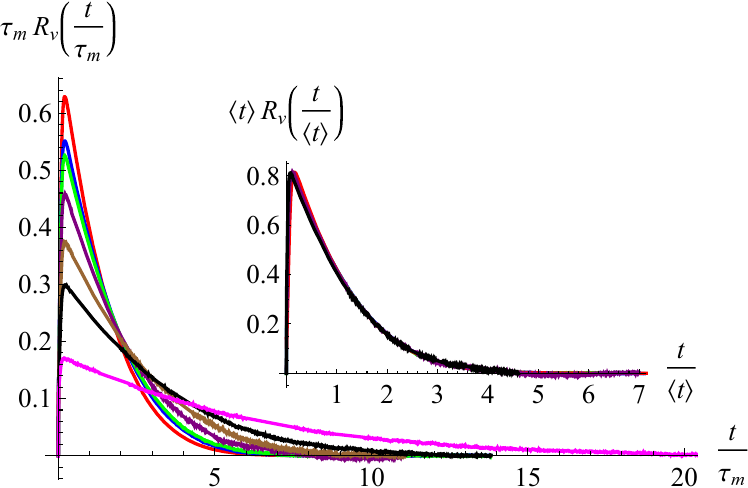}
  \caption{Main plot: The response function rescaled with the theoretically predicted time scale $\tau_m$. Velocities from top to bottom are $v=0$ (red), $v=0.3$ (blue), $v=0.5$ (green), $v=1$ (violet), $v=1.5$ (brown), $v = 2$ (black) and $v =3\approx f_{\rm c}$ (magenta). Inset: Scaling collapse for the velocities $v \leq 2$, using the measured time scale $\left<t \right>$ instead of $\tau_m$ for the rescaling.  (DNS, $\delta u = 10^{-6}$, $\delta t = 10^{-4}$,   $N=10^{7}$).}
\label{FigvelResponsescalingCollapseM2}
\end{figure}
The response function at a finite driving velocity  and the associated time scales are defined as 
\bea\label{Rv}
R_v(t) &:=& \lim_{\delta w\to 0} \frac{1}{\delta w} \left[\left< \dot u(t) \right>^{\delta w}_v - v \right],\\
\label{tau(v)}
\left< t\right>_v &=& \int_0^\infty \rmd t  \, R_v(t) t.
\eea
The expectation value on the r.h.s.\ of \Eq{Rv} is the expectation of $\dot u(t)$, given a driving velocity $v$, and an additional kick at time $t=0$; we have subtracted the expectation without a kick, $\left< \dot u(t) \right>^{\delta w=0}_v=v$.
The time scale $\left< t\right>_v$ defined in \Eq{tau(v)} should then behave as 
\be\label{tv}
\frac1{
\left< t\right>_v}\simeq f_{\rm c} -v, \qquad v\mbox{ small.}
\ee
What appears on the r.h.s.\ is the relative velocity between the ``ballistically'' moving particle as given by   \Eq{P-udot-DPM} and the advancement of the confining potential.  
On the other hand, at large driving velocity, the disorder   acts as a thermal noise, and the response function   reduces to that of the free theory, equivalent to that of the ABBM model, resulting in 
\be
\frac1{
\left< t\right>_v}\simeq  \tau_{\rm ABBM}^{-1} = m^2, \qquad  v\mbox{ large.}
\ee
These two curves are plotted on Fig.~\ref{figTimescale}. They intersect at $v=f_{\rm c}-m^2$. Intuitively we expect the transition to take place at $v=f_{\rm c}$, and to be smoothed out due to the finite width of the velocity distribution \eq{P-of-udot}. As we can see on Fig.~\ref{figTimescale},  the transition gets sharper  when $m^2$ decreases. A phase transition at $v=f_{\rm c}$ is expected in the limit of $m^2\to 0$. 

Our final observation is that
the time scale $\left<t\right>_v$ defined in \Eq{tv} can   be used for a scaling collapse of the response functions $R_v(t)$ defined in \Eq{Rv}, see Fig.~\ref{FigvelResponsescalingCollapseM2}. 
Details of the  analysis are given in Ref.~\cite{terBurgPhD}.

\subsection{Avalanche-size distribution}
\label{s:Avalanche-size distribution}

The avalanche-size distribution is shown on  Fig.~\ref{f:1}, for $m^2=10^{-4}$, overlayed with two theoretical curves: the kicked ABBM model as given in \Eq{PwS(S)=ABBM} (bright dotted cyan), and the DPM given in \Eq{PofS-discrete} (dark blue dashed).

For small $S$, one sees the $S^{- 3/2}$ behavior characteristic for ABBM coincide with the theoretical curve \eq{PwS(S)=ABBM} (cyan, dashed). For large $S$, this crosses over to the prediction of \Eq{PofS-discrete} for the DPM with avalanche exponent $\tau=0$  (blue, dashed). Note that there are no adjustable parameters, and both curves respect their  own normalization. The crossover takes place at $S\approx 1$,  
and extends over about half a decade in both directions. 

\subsection{Avalanche-duration distribution}
\begin{figure}
\Fig{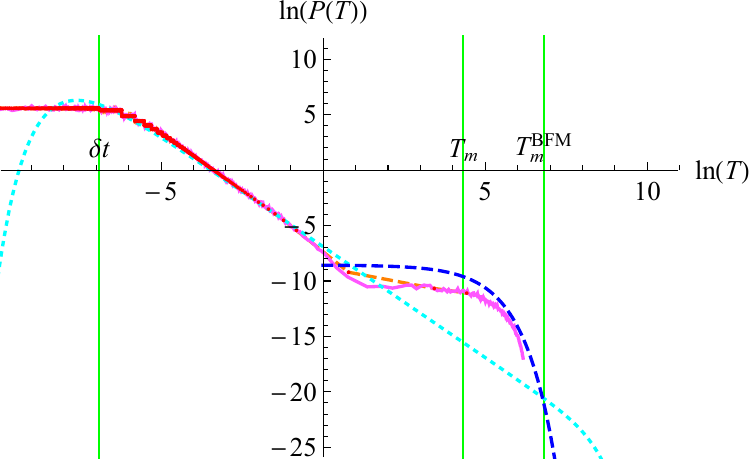}
\caption{$P(T)$ measured numerically via INS for $m^{2}=10^{-3}$, ($\delta w = 1$, $\delta t=10^{-3}$, $N= 5\times 10^8$). This is compared to the  theory prediction   \eq{P-duration1} for ABBM  (cyan dotted), and \eq{P(T)DPM} for DPM (blue, dashed).}
\label{f:P(T)simul}
\end{figure}
The avalanche-duration distribution is shown on Fig.~\ref{f:P(T)simul}, superimposed with the two theoretical predictions \eq{P-duration1} for ABBM  (cyan dotted), and \eq{P(T)DPM} for the DPM (blue, dashed). The simulations agree very well with ABBM for small times, while the overall weight for large times is seemingly overestimated in \Eq{P(T)DPM}, contrary to what one saw on Fig.~\ref{f:1} for the avalanche-size distribution. Normalization issues are more pronounced since the singularity for small times $P(T)\sim T^{-2}$ is stronger than the $P(S)\sim S^{-3/2}$ for the avalanche size.  Another reason for the slow convergence is the relatively broad avalanche-velocity distribution, assumed to be restricted to $\dot u=f_{\rm c}$ in the derivation of \Eq{P-duration1}.

\subsection{Correlators $\Delta_v(w)$, $\Delta_{\dot u}(w)$, $\Delta_F(w)$}
\label{s:3 Deltas}

\begin{figure}[t]
\includegraphics[width=\columnwidth]{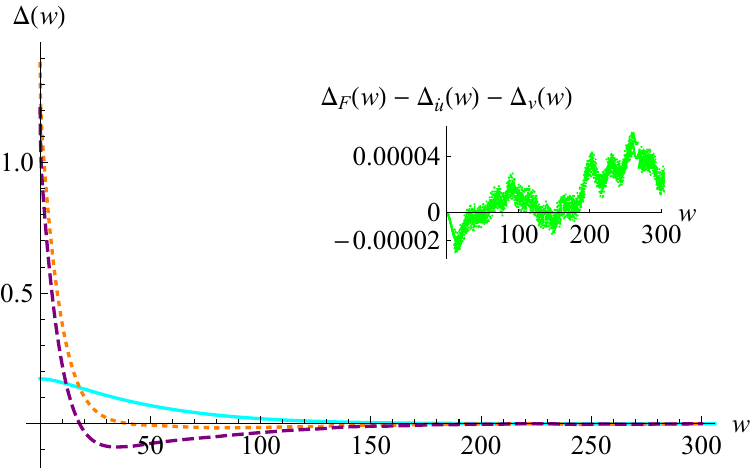}
   \caption{The measured $\Delta_v$ (cyan) , $\Delta_F$ (orange, dashed)  and $\Delta_{\dot{u}}$ (purple, dotted) for $m^2 = 0.01$ at $v = 0.5$. The inset shows that the difference $\Delta_F - \Delta_v - \Delta_{\dot{u}}$ vanishes.   (DNS, $\delta u =0.01$, $\delta t = 10^{-4}$, $N = 10^{8}$).}
\label{FigDeltaFDeltaUdotDeltaURelations}
\end{figure}

In section \ref{s:Other auto-correlation functions}, we   defined the three correlation functions $\Delta_v(w)$, $\Delta_{\dot u}(w)$, and $ \Delta_{  F}(w)$, which are the correlations of three forces: confining well, friction, and disorder. 
Since they add up to zero, two variables are independent, and one would expect three independent correlation functions, which are most symmetrically expressed as  $\Delta_v(w)$, $\Delta_{\dot u}(w)$, and $ \Delta_{  F}(w)$.
This expectation is incorrect: there are only two independent quantities, summarized in the relation
\be\label{Delta-relation}
\Delta_v(w)+\Delta_{\dot u}(w) = \Delta_{  F}(w).
\ee
A numerical verification is presented on Fig.~\ref{FigDeltaFDeltaUdotDeltaURelations}. An analytical proof can be given too: identifying $w \equiv vt$, one has 
\bea
&&\Delta_F(w-w') = \left< F_w F_{w'}\right>^{\rm c} \nn\\
&&~~=\left< [\dot u_w+m^2(u_w-w)][\dot u_{w'}+m^2(u_{w'}-{w'})]\right>^{\rm c}\qquad\nn\\
&&~~=\left<  \dot u_w \dot u_{w'} \right>^{\rm c}+ m^4 \left<  (u_w-w) (u_{w'}-{w'})\right>^{\rm c}\nn\\
&&~~~~~~+m^2\left<\dot u_w (u_{w'}-{w'})+  (u_w-w) \dot u_w  \right>^{\rm c}.
\eea
The last term can be written as 
\bea
&&\left<\dot u_w (u_{w'}-{w'})+  (u_w-w) \dot u_w  \right>^{\rm c} \nn\\
&& = (\partial_t +\partial_t')\left<(u_w-w) (u_{w'}-{w'})\right>^{\rm c} \nn\\
&& = \frac 1 v \left( \partial_w + \partial_w'\right) \Delta_v(w-w') = 0.
\eea
This proves \Eq{Delta-relation}.

\begin{figure}[b]
\Fig{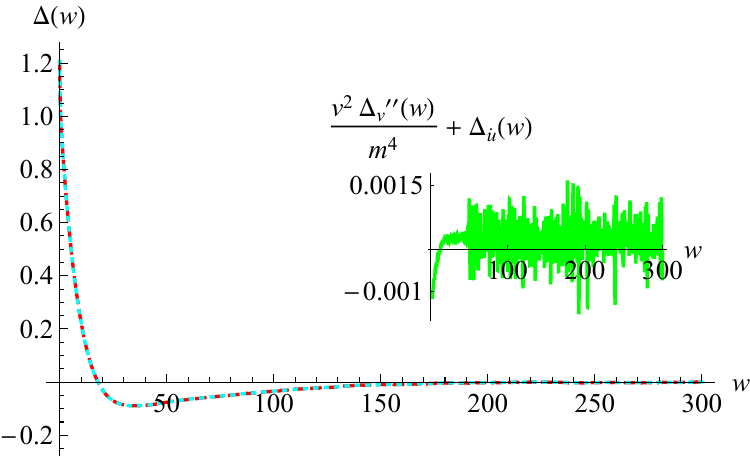}
\caption{Numerical test of \Eq{Deltadotu=-Delta''}  for $m^2 =10^{-2}$, $v = 0.5$. Plotted is $\Delta_{\dot u}(w)$ (cyan, dottedd) and $ -\frac{v^{2}}{m^{4}} \Delta''_{v}(w)$ (red, dashed). The inset shows the difference.    (DNS, $\delta u =0.01$, $\delta t = 10^{-4}$, $N = 10^{8}$).}
\label{f:Deltadotu=-Delta''}
\end{figure}

On Fig.~\ref{f:Deltadotu=-Delta''}, 
we find numerically satisfied another relation,
\be
m^4 \Delta_{\dot u}(w) =  -v^2 \partial_w^2 \Delta_v(w)\ .
\ee
This relation expresses the time derivative of $u(t)= u_w$ by its dependence on $w=vt$: $\partial_t u(t) = v\partial_w u_w$.
Thus knowing $\Delta_{v}(w)$ is enough, and the two other quantities can be expressed as
\bea
\label{Deltadotu=-Delta''}
\Delta_{\dot  u}(w) &=& -\frac{v^{2}}{m^{4}} \Delta''_{v}(w) ,\\
\Delta_{F}(w) &=& \Delta_{v}(w) -\frac{v^{2}}{m^{4}} \Delta''_{v}(w).
\eea
Knowing $\Delta_v(w)$ it is easy to find $\Delta_{\dot u}(w)$ by taking two derivatives. It is already less obvious to reconstruct 
$\Delta_v(w)$ from $\Delta_{\dot u}(w)$, as this procedure involves two integration constants. In principle the latter are fixed since all correlations vanish  for $w\to \infty$; in practice, however,  fluctuations grow   with increasing  $w$, and they show up in the two integration constants. 
It is even more involved to reconstruct $\Delta_v(w)$ from $\Delta_F(w)$: Formally, this can be achieved by the series
\be
\Delta_v(w) = \sum_{n=0}^\infty \left[\frac{v}{m^2} \frac{\rmd } {\rmd w}\right]^{2n} \Delta_F(w).
\ee
As it is  rather badly converging, it may well be useless in practice.

\subsection{Measuring  $\Delta(w)$ at vanishing driving velocity}
\label{s:Measuring Delta(w) at vanishing driving velocity}
\begin{figure}
\Fig{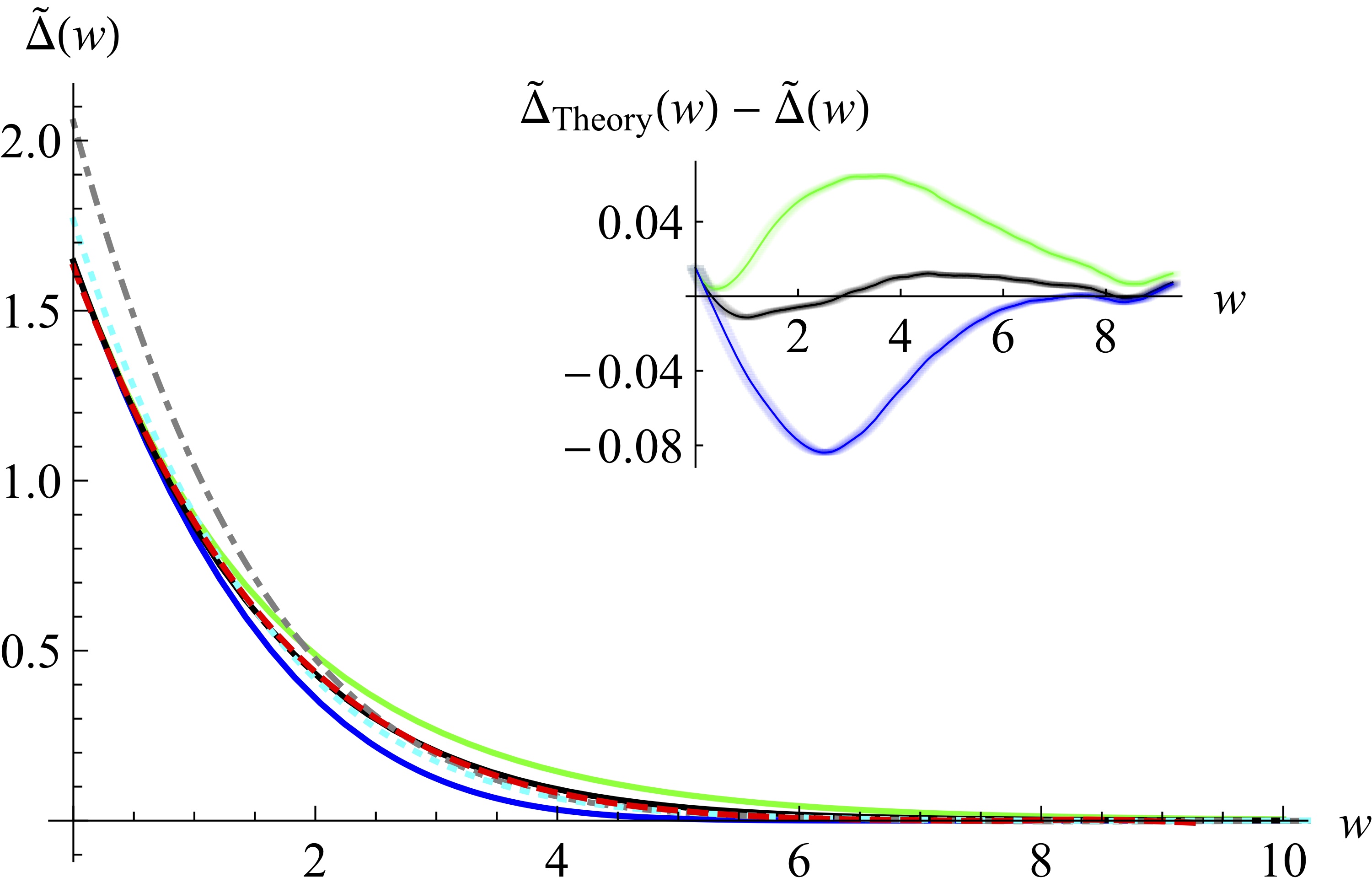}
\caption{Convergence of the measured $\tilde \Delta(w)$ for $m^2=10^{-2}$ (gray, dot-dashed),  $m^2=10^{-3}$ (cyan, dotted) and  $m^2=10^{-4}$ (red,  dashed). This is compared to three theoretical curves:  an exponential function with the same slope at $w=0$ (green, top curve), \Eq{84} (black, middle curve), and the 1-loop FRG-result (blue, bottom curve), see Eqs.~(4.5) and (4.11) of \cite{LeDoussalWieseChauve2002}.
The inset shows the difference of the measured $\tilde \Delta(w) $ at $m^2=10^{-4}$, and the theoretical curves in their respective colors, together with error bars at a 99$\%$ confidence level. (DNS, $\delta u = 0.01$, $\delta t = 10^{-4}$, $N = 10^8$, $\delta w = 0.1$).}
\label{f:18}
\end{figure}
In an experiment, it is difficult to measure $\Delta(w)=\lim_{v\to 0}\Delta_v(w)$. In our simulation, we can do this: We move the parabola from $w \to w+\delta w$, and then wait until the dynamics cedes. 
Due to Middleton's theorem \cite{Middleton1992}, the position $u_w$ is history independent.  
 From $u_w$ we  obtain $\Delta(w)$ via formula \eq{16}. 
The DPM \cite{LeDoussalWiese2008a}   predicts that 
\bea\label{83}
 \Delta(w) &=&  m^4 \rho_m^2  \tilde \Delta (w /\rho_m), \\ 
 \label{84}
 \tilde \Delta(w) &=&   \text{Li}_2(1 - \rme^{|w| }) + \frac{w^2}{ 2}+ \frac{\pi^2}{6}  .
 \eea
To check this relation, we study the experimentally measured $\Delta_{\rm exp}(w)$, and then invert \Eq{83} to obtain
\be\label{85}
\tilde \Delta_{\rm exp}(w) =  \frac1{m^4 \rho_m^2}   \Delta_{\rm exp}(w/\rho_m).
\ee 
Figure \ref{f:18} shows convergence against the theoretical result.

\begin{figure}
\Fig{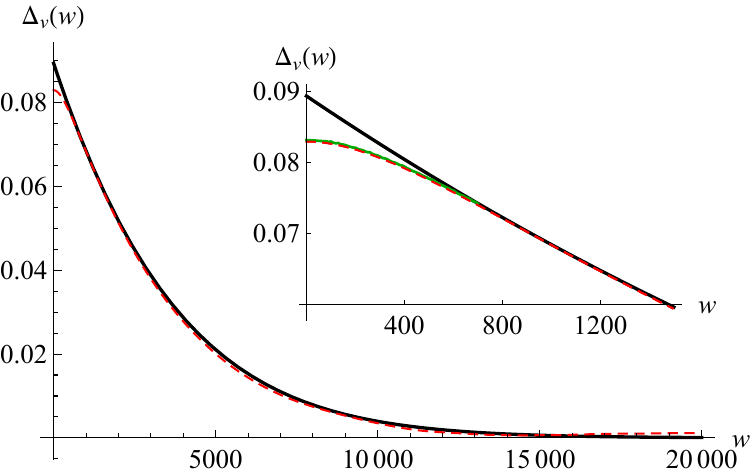}
\caption{The numerically measured  $ \Delta_v(w)$ for $v=0.5$  (red, dashed, with $95\%$-confidence intervals in shaded red) as defined in \Eq{De-v}.  For $w>1000$ it agrees well with  the theoretical prediction \eq{83} (black, solid). The inset shows   the same curves, superimposed with the theoretical prediction \eq{83}
 folded according to \Eq{Delta-u-theory} with the   numerically measured response function (green), lying underneath $\Delta_v(w)$ obtained numerically (red, dashed). (DNS, $\delta u = 0.01$, $\delta t = 10^{-4}$, $N = 10^7$).}
\label{f:tilde-Delta-exp-check}
\end{figure}

\subsection{The disorder correlator $\Delta_v(w)$ at a finite driving velocity}
\label{s:The disorder correlator Delta_v(w) at a finite driving velocity}
For a finite driving velocity, \Eq{Delta-u-theory} predicts that the measured $\Delta_v(w)$ is obtained from the zero-velocity correlator $\Delta(w)$ by folding with two response functions. In the inset of  Fig.~\ref{f:tilde-Delta-exp-check}  we check  that this construction works for $ v=0.5 $, by folding the numerically measured zero-velocity correlator $\Delta(w)$ with the numerically measured  zero-velocity response function $R(t)$.

\begin{table*}[t]
\begin{tabular}{ |p{2.9cm}|p{2.cm}||p{3cm}|p{2.5cm}|  p{2cm}| }
\hline
parameters&  $\tau := \int_t t R(t) $ & $\tau$ from boundary layer analysis via \Eq{BL} &$\tau$ from differential  \Eq{242} & $\tau$ from \Eq{vtau-estimate}  \\
 \hline
$m^2 = 0.01, v= 0.1 $   & 14.08    &14.10&   14.08 &14.87 \\
$m^2 = 0.01, v= 0.3 $  & 14.08    &14.08&   14.08 & 13.90 \\
 $m^2 = 10^{-4}, v= 0.5 $  & 596  &587&   587 & 664 \\ 
 \hline
\end{tabular}
\caption{Comparison of the measured time scale $\tau$, and its estimation using the methods of section \ref{s:Boundary-layer analysis}-\ref{s:Differential equation}. The estimation via the differential equation \eq{BL} is the most precise, followed by the boundary-layer analysis using \Eq{242}, whereas the  approximation \eq{vtau-estimate} has a relatively large error. }
\label{tau-table}
\end{table*}

We   numerically established that the response function evolves with $v$ from their zero-velocity  
limit (see Fig.~\ref{FigvelResponsescalingCollapseM2}), and   the same may  happen for the effective disorder-correlator, technically the effective action of the theory. While the agreement shown in Fig.~\ref{f:tilde-Delta-exp-check}   should be sufficient for a typical experiment, the question arises what happens when we increase the driving velocity. Intuitively, as for the response function, we expect a transition when the driving velocity exceeds $f_{\rm c}$.

\begin{figure}[t]  
\Fig{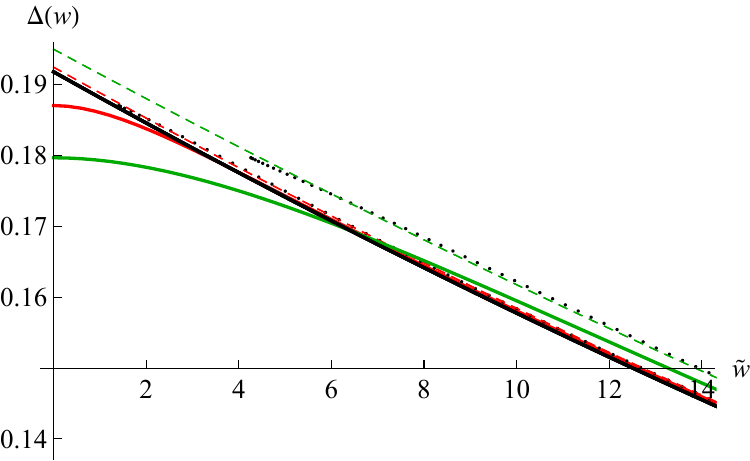}
\caption{Measurement of $\Delta_v(w)$ for $v=0$ (solid blue, top), $v=0.1$ (solid red, middle), and $v=0.3$ (solid green, bottom).
We show unfolding via a boundary layer analysis, using $\tau :=\left< t\right>= 14.08$ as measured for the response function, both for $v=0.1$ (black dots), prolonged to zero (red, dashed) and $v=0.3$ (black dots), prolonged to zero (green, dashed). The variable $\tilde w= \sqrt{w^2 +(\tau v)^2}$.
 (DNS, $\delta t=10^{-4}$, $\delta u=10^{-2}$, $N=10^8$.)
}
\label{FigCheckUnfolding1}
\end{figure}

\subsection{Reconstructing $\Delta(w)$ from  $\Delta_v(w)$ }
\label{s:reconstruction-simul}
In sections \ref{s:Boundary-layer analysis} to \ref{s:Differential equation} we proposed two procedures to reconstruct the zero-velocity correlator $\Delta(w)$, based either on a boundary-layer analysis (section \ref{s:Boundary-layer analysis}), or a differential equation (section \ref{s:Differential equation}). These methods contain a parameter, the time-scale $\tau$ of the response function, in the combination $\delta_w=\tau v$, and allow us to extract the latter if unknown.
Let us analyze these methods in turn.

\subsubsection{Boundary-layer analysis}
\label{s:Boundary-layer analysis-sim}
As proposed in section \ref{s:Boundary-layer analysis},  we use the boundary-layer formula \eq{BL} to plot on Fig.~\ref{FigCheckUnfolding1}
the measured $\Delta_v(w)$ against $\tilde w=\sqrt{w^2+ \delta _w ^2}$,  $\delta_ w=v\tau$, using   the timescale $\tau=\left< t\right> $ measured from the response function.
By extrapolation to $\tilde w=0$ we obtain the full $\Delta(w)$. This works decently well, especially for the smaller driving velocity $v=0.1$.

\subsubsection{Differential equation}
\begin{figure}[t]   
\Fig{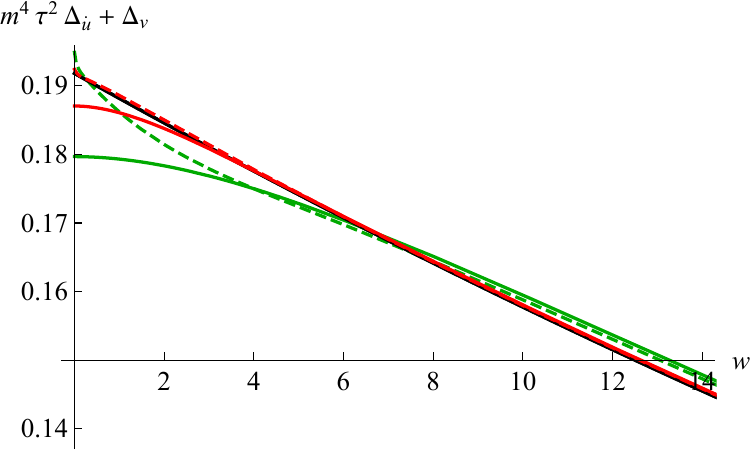}
  \caption{Measurement of $\Delta_v(w)$ for $v=0$ (solid blue, top), $v=0.1$ (solid red, middle), and $v=0.3$ (solid green, bottom).
We show unfolding via the differential equation \eq{242} using $\tau :=\left< t\right>= 14.08$ as measured for the response function, both for $v=0.1$ (red, dashed) and $v=0.3$ (green, dotted). (DNS, $\delta t=10^{-4}$, $\delta u=10^{-2}$, $N=10^8$.)}
\label{FigCheckUnfolding2}
\end{figure}

A second unfolding procedure was proposed in section \ref{s:Differential equation}.  Repeating the analysis performed on Fig.~\ref{FigCheckUnfolding1}, we  show  on Fig.~\ref{FigCheckUnfolding2} the reconstructed $\Delta(w)$. 
For $v=0.1$ the unfolding procedure reproduces the $v=0$ correlator with high precision. The 
agreement is not as good for $v=0.3$.

\subsubsection{Reconstructing the time scale $\tau$}
In the last two sections, we   tested the reconstruction procedures under the assumption that  the   time scale $\tau$ was known from an independent  measurement of the response function. In practice, $\tau $ may not be available. 
In that case, we   try to find    $\delta_ w=v\tau$, 
 which best removes the curvature of $\Delta_v(w)$. The estimated values for $ \tau$ are reported on table \ref{tau-table}, together with an estimate of $\tau$ from \Eq{vtau-estimate}. 
While the latter gives only fair results, both reconstruction procedures allow us to extract the time scale $\tau$ rather precisely.

\section{Summary and conclusion}
We have shown that there is more to mean-field theory than replacing the effective force landscape by a random walk (ABBM model). As forces cannot grow unboundedly, they must finally saturate, leading to a different regime of uncorrelated forces (DPM). Above we propose to describe the  crossover by modeling forces as an Ornstein-Uhlenbeck process. Using numerical simulations supported by   analytical results for all key ingredients, we  quantify the signatures   expected in experiments. The key observable is the effective force-force correlator, which is readily accessible in experiments. As experiments necessitate a finite driving velocity, the measured signal is always smeared out. Above we developed and tested  procedures to reconstruct   the zero-velocity response  and force-force correlations from finite-velocity measurements.  We have already tested our procedure with success for magnetic domain walls (collaboration with G.~Durin) and  knits (collaboration with A.~Douin and F.~Lechenault). Each of these systems has its own peculiarities,   on which we will report   in forthcoming publications. 

\acknowledgements

It is a pleasure to thank A.~Kolton, G.~Mukerjee, A.~Rosso and B.~Walter   for fruitful discussions. The questions raised here are inspired by our ongoing collaboration with experimentalists  and we are grateful for the inspirations brought to us by  G.~Durin (magnetic domain walls), as well as  A.~Douin and F.~Lechenault (knitting).

\appendix

\section{Correlations of an Ornstein Uhlenbck process}
\label{a:OU-correlations}
Suppose that \be\label{Ornstein-Uhlenbeck-bis}
 \partial_u F \big(u  \big) =  -  F(u)+ \xi(u) .
\ee
This equation is solved by 
\be
F(u) = \int_{-\infty}^u \rmd u_1 \, \rme^{- (u-u_1)}\xi(u_1) .
\ee
It leads to microscopic correlations 
\bea
&&\Delta(u-u') := \overline {F(u) F(u')} ~~~~~~~\qquad\qquad\qquad\qquad~~~\nn  \\
&&= \int_{-\infty}^u \rmd u_1 \int_{-\infty}^{u'} \rmd u_2\, \rme^{- (u+u' -u_1-u_2)} \overline{\xi(u_1)\xi(u_2)}
\nn
\\
&&= 2   \int_{-\infty}^{{\rm min}(u,u')} \rmd \tilde u \,\rme^{- (u+u' - 2\tilde u)}  \nn
\\
&&=  \rme^{- |u-u'|}.
\eea

\section{Numerical implementations}
\label{s:Numerical implementations}
We used several numerical implementations:
\begin{enumerate}
\item[(i)] {\em Direct numerical simulation (DNS)}.
To solve the coupled set of differential equations  \eq{EOM}--\eq{220}
we  use a space discretization $\delta u =10^{-6}$ to $10^{-2}$ (depending on $m$) to first obtain  the random forces $F(u)$ for $u=n \delta u$, $n\in \mathbb N$. $F(u)$  is then linearly interpolated  between these  points.  We finally  solve \Eq{EOM} with the Euler method, using a time-discretization of $\delta t = 10^{-4}$.  
\item[(ii)] {\em Improved numerical solver (INS)}. In the  coupled equations of motion \eq{EOM}-\eq{220}, the force $\ca F(t):= F\big(u(t)\big) $
can statistically equivalently  be modeled as  \cite{DornicChateMunoz2005,DobrinevskiPhD,KoltonLeDoussalWiese2019} 
\bea\label{mult-noise}
\partial_ t \ca F(t) &=& - \ca F(t) + \sqrt{\dot u(t)} \eta(t) , \\
\left< \eta(t) \eta(t') \right> &=& 2 \delta(t).
\eea
The advantage of this scheme is that there is  only one parameter, namely   the time discretization $\delta t$, but none for the space discretization   $\delta u$. The effective space discretization is $\delta u \approx \dot u(t) \, \delta t$, thus is   finer when the system moves slowly. Treating the multiplicative noise of \Eq{mult-noise} however demands some care \cite{Munoz2004}. We use both a direct simulation with a very small time-step as the scheme prosed in \cite{DornicChateMunoz2005} and tested in \cite{DobrinevskiPhD,KoltonLeDoussalWiese2019}.

\end{enumerate}
The number of samples is denoted 
\be
N:=\mbox{number of samples}.
\ee
Most of our results were verified with both schemes, DNS and INS.


\ifx\doi\undefined
\providecommand{\doi}[2]{\href{http://dx.doi.org/#1}{#2}}
\else
\renewcommand{\doi}[2]{\href{http://dx.doi.org/#1}{#2}}
\fi
\providecommand{\link}[2]{\href{#1}{#2}}
\providecommand{\arxiv}[1]{\href{http://arxiv.org/abs/#1}{#1}}
\providecommand{\hal}[1]{\href{https://hal.archives-ouvertes.fr/hal-#1}{hal-#1}}
\providecommand{\mrnumber}[1]{\href{https://mathscinet.ams.org/mathscinet/search/publdoc.html?pg1=MR&s1=#1&loc=fromreflist}{MR#1}}

\tableofcontents

\end{document}